\newcommand{\version}{July 5, 2000}

\documentclass[12pt]{article}
\usepackage{a4,amsthm,amsfonts,latexsym,amssymb}

{\catcode `\@=11 \global\let\AddToReset=\@addtoreset}
\AddToReset{equation}{section}

\swapnumbers
\theoremstyle{plain}
\newtheorem{thm}{THEOREM}[section]
\newtheorem{cl}[thm]{COROLLARY}
\newtheorem{lem}[thm]{LEMMA}
\newtheorem{proposition}[thm]{PROPOSITION}
\theoremstyle{definition}
\newtheorem{rem}[thm]{Remark}

\newcommand{\beq}{\begin{equation}}
\newcommand{\eeq}{\end{equation}}
\def\beqa{\begin{eqnarray}}
\def\eeqa{\end{eqnarray}}

\newcommand{\infspec}{{\rm inf\ spec\ }}
\newcommand{\nab}{\left(-i\nabla+\beta\A(\x)\right)}
\newcommand{\R}{{\mathbb R}}

\newcommand{\Ll}{{\mathcal L}}
\newcommand{\Hh}{{\mathcal H}}
\newcommand{\Cc}{{\mathcal C}}
\newcommand{\eps}{\varepsilon}
\newcommand{\A}{{\bf a}}
\newcommand{\abf}{{\bf a}}
\newcommand{\B}{{\bf b}}

\newcommand{\x}{{\bf x}}
\newcommand{\y}{{\bf y}}

\newcommand{\0}{{\bf 0}}
\newcommand{\rr}{{\bf r}}

\newcommand{\xperp}{\x^\perp}
\newcommand{\yperp}{\y^\perp}

\newcommand{\xpar}{x^\parallel}
\newcommand{\ypar}{y^\parallel}
\newcommand{\ppar}{p^\parallel}
\newcommand{\Lpar}{L^\parallel}
\newcommand{\Tr}{{\rm Tr}}
\newcommand{\half}{\mbox{$\frac{1}{2}$}}
\newcommand{\rg}{\rho_\Gamma}

\newcommand{\rh}{\rho_{\Gamma^{\rm H}}}
\newcommand{\dx}{\frac\partial{\partial x}}
\newcommand{\dy}{\frac\partial{\partial y}}
\newcommand{\E}{{\mathcal E}^{\rm MH}}

\newcommand{\tE}{{\mathcal E}^{\rm MH}_{\beta , \rm lin}}
\newcommand{\Edens}{{\hat{\mathcal E}}^{\rm MH}}
\newcommand{\uw}{\underline w}

\date{\small\version}

\begin{document}
\markboth{\scriptsize{BS \version}}{\scriptsize{BS \version}}
\title{\bf{Atoms with bosonic ``electrons'' in \\ strong magnetic fields}}
\author{\vspace{5pt} Bernhard Baumgartner$^1$ and Robert Seiringer$^2$\\
\vspace{-4pt}\small{Institut f\"ur Theoretische Physik, Universit\"at Wien}\\
\small{Boltzmanngasse 5, A-1090 Vienna, Austria}}

\maketitle

\begin{abstract}
We study the ground state properties of an atom with nuclear
charge $Z$ and $N$ bosonic \lq\lq electrons\rq\rq\ in the presence
of a homogeneous magnetic field $B$. We investigate the mean field
limit $N\to\infty$ with $N/Z$ fixed, and identify three different
asymptotic regions, according to $B\ll Z^2$, $B\sim Z^2$, and
$B\gg Z^2$. In Region 1 standard Hartree theory is applicable.
Region 3 is described by a one-dimensional functional, which is
identical to the so-called Hyper-Strong
functional introduced by Lieb, Solovej and Yngvason 
for atoms with fermionic electrons in the
region $B\gg Z^3$; i.e., for very strong magnetic fields the ground state
properties of atoms are independent of statistics. For Region 2  
we introduce a general {\it magnetic
Hartree functional}, which is studied in detail. It is shown that
in the special case of an atom it can be restricted to the subspace of zero
angular momentum parallel to the magnetic field, which simplifies
the theory considerably. The functional reproduces the energy and
the one-particle reduced density matrix for the full $N$-particle
ground state to leading order in $N$, and it implies the
description of the other regions as limiting cases.
\end{abstract}

\footnotetext[1]{E-Mail:
\texttt{baumgart@ap.univie.ac.at}}
\footnotetext[2]{E-Mail:
\texttt{rseiring@ap.univie.ac.at}}

\newpage

\tableofcontents

\section{Introduction}

The ground states of atoms with many electrons in magnetic fields
have been studied in \cite{LSY94a, LSY94b, BSY00}, and their
energies have been evaluated, exactly to leading order, as some of
the physical parameters tend to infinity. The atoms have been
modeled by the nonrelativistic quantum mechanics of $N$ fermionic
electrons, with an unmovable pointlike nucleus of charge $Z$ in a
homogeneous magnetic field of strength $B$. In order to shed some
more light onto the interplay of the involved laws of physics, we
investigate the effects of changing one of them: What would
happen, if the electrons were bosons?

So we study the ground state of the Hamiltonian - written in
appropriate units -

\beq\label{ham}
\underline H_{N,Z,B}=\sum_{i=1}^N\left( H_{B,i}-B-
\frac{Z}{|\x_i|}\right)+ \sum_{i<j}\frac{1}{|\x_i-\x_j|},
\eeq
where we set
\beq
H_{B,j}=\left(-i\nabla_j +B \abf(\x_j)\right)^2.
\eeq
The vector potential is given by $\abf(\x)={\bf e}^\parallel
\times \x/2$, where $\x\in\R^3$, ${\bf e}^\parallel$ is the unit
vector parallel to the magnetic field in $z$-direction. This
Hamiltonian acts on the {\it symmetric} subspace of
$\Ll^2(\R^{3N},d^{3N}\x)$. We subtract $B$ for every particle
because we are interested in the binding energy, which is now
equal to the ground state energy $\underline E(N,Z,B)=\infspec
\underline H_{N,Z,B}$.

\bigskip
In the {\bf study of asymptotics}, as $B$ and $Z$ tend to
infinity, we find a division into {\it three different regions}.
They are - always in appropriate units - characterized by: $B \ll
Z^2$, $B  \sim  Z^2$, $B \gg Z^2$. This is in contrast to atoms
with fermionic electrons, where five different regions have been
identified: $B \ll Z^{4/3}$, $B \sim Z^{4/3}$, $Z^{4/3} \ll B \ll
Z^3$, $B \sim  Z^3$, $B \gg Z^3$. See \cite{LSY94a} and references
therein.

A simple heuristic argument: The length scale, which is typical
for the quantum effects of a {\it single} particle in the magnetic
field, is $ \sim B^{-1/2}$. Typical energies are the differences
of the Landau levels, $2B$. On the other hand, the length scale
typical for a particle in the Coulomb potential only is $ \sim
Z^{-1}$ and hence the typical energy range is $ \sim Z^2$. In
Region 2, where $B \sim Z^2$, the magnetic and the Coulombic
effects are therefore of the same order of magnitude. In Region 1,
where $B \ll Z^2$, the Coulomb effects dominate in all directions.
Magnetic effects will not contribute in leading order. In Region
3, where $B \gg Z^2$, the magnetic effects dominate the dynamics
perpendicular to the magnetic field. The electron with low energy
is confined to the lowest Landau band, the typical wave functions
are squeezed to needles with diameter $ \sim B^{-1/2}$. (See
\cite{AHS81, FW94} for a detailed rigorous treatment and for
citations concerning the history of this problem.) Turning from
the one-body system to the $N$-body problem, we remark that Bose
statistics has no effect on the size of the ground states, if the
pair interactions are ignored. Moreover, the repulsion of the
particles is of the same order of magnitude as the attraction of
the nucleus, if $N\sim Z$. So the scaling properties of the
lengths and energies per particle remain the same, and the
distinction of the three regions for many bosons is the same as
for a single electron.

We remark, that we can moreover identify a Region 4 with strong
magnetic field, where the nuclear charge $Z$ is fixed. In this
region the asymptotics is, in leading order, independent of the
statistics, as has been noted in its evaluation in \cite{BSY00}.
It is to be described by the model of one-dimensional atoms with
delta-function interactions.

{\bf Scaling:} In the following we will use the parameters
$\lambda=N/Z$, $\beta=B/Z^2$ besides $N$. By scaling
$\x\rightarrow\x/Z$ the operator $Z^{-2}\underline H_{N,Z,B}$ is
unitarily equivalent to
\beq\label{ham2}
H_{N,\lambda,\beta}=\sum_{i=1}^N\left(H_{\beta,i}-\beta-
\frac{1}{|\x_i|}\right)+\frac\lambda N
\sum_{i<j}\frac{1}{|\x_i-\x_j|},
\eeq
with ground state energy $E(N,\lambda,\beta)=Z^{-2}\underline
E(N,Z,B)$. We are interested in the limit $N\to\infty$ of
$N^{-1}E(N,\lambda,\beta)$ with $\lambda$ fixed. In Region 1 this
limit is coupled with $\beta\to 0$, and in Region 3 with
$\beta\to\infty$, while $\beta$ is fixed in Region 2. The
asymptotics of the atomic structure and of its energy for large
$N$ in the three regions is modeled by energy functionals in
generalized Hartree theory.
\bigskip

In {\bf Region 1} standard Hartree theory is applicable. The
energy functional, a functional of the density $\rho$, is
\beq\label{hartree1}
\mathcal{E}^{\rm H}[\rho ]=\int |\nabla \rho ^{1/2}(\x)|^{2}d^3\x
-\int \frac{1}{|\x|}\rho (\x)d^3\x +D[\rho ,\rho ],
\eeq
where
\beq\label{drho}
D[\rho,\rho]=\frac 12\int\frac{\rho(\x)\rho(\y)}{|\x-\y|}d^3\x
d^3\y .
\eeq
Its ground state energy is
\beq\label{nrg1a}
E^{\rm H}(\lambda)=\inf_{\rho , \hspace{4pt} \int \rho
=\lambda}\mathcal{E}^{\rm H}[\rho ].
\eeq
It is known, \cite{BL83}, that
\beq
\lim_{N\to\infty}\frac{1}{N}E(N,\lambda,0)=
\frac{1}{\lambda}E^{\rm H}(\lambda).
\eeq
We will extend this result to
\begin{thm}[Energy asymptotics for Region 1]\label{nrg1}
If $N\to \infty$ and \newline $\beta =\beta (N) \to 0$ with
$\lambda$ fixed, then
\beq
\lim_{N\to\infty}\frac{1}{N}E(N,\lambda,\beta)
=\frac{1}{\lambda}E^{\rm H}(\lambda).
\eeq
\end{thm}
\bigskip

In {\bf Region 2} Hartree theory has to be generalized. The basic
idea remains: the electrons occupy the ground states of an
effective one-particle Hamiltonian with a mean field potential
which has to be determined by self-consistency. Now in the
presence of a magnetic field, the ground state of the effective
Hamiltonian may a priori be degenerate, so that the electrons can
be distributed over a larger set of states. To take this into
account, one has to consider in general one-particle {\it density
matrices} $\Gamma$ in this {\it Magnetic Hartree Theory}. The
energy functional is
\beq\label{hartree2}
\E_{\beta}[\Gamma]=\Tr[(H_\beta-\beta)\Gamma]-
\int\frac{1}{|\x|}\rg(\x)d^3\x +D[\rg,\rg],
\eeq
where $\rg(\x)$ is the density defined by $\Gamma$, with $\int \rg
= \Tr [\Gamma]$. We define the Hartree energy $E^{\rm
MH}(\lambda,\beta)$ as
\beq\label{nrg2a}
E^{\rm MH}(\lambda,\beta)=\inf_{\Gamma , \hspace{4pt}
\Tr[\Gamma]=\lambda}\E_{\beta}[\Gamma].
\eeq

This general form of magnetic Hartree theory is necessary, if some
extra external potential is added, or if one considers, e.g.,
molecules. For, in the presence of magnetic fields, some well known
facts of ordinary quantum mechanics are no longer true: the ground
state may be degenerate, and, in the case of an axially symmetric
system, it can happen that the energy is not minimized by states with
zero angular momentum \cite{LO77, AHS78}. But for the atom 
without perturbing forces, it turns
out that the theory can be reduced to the consideration of pure
states, rank one density matrices, with a zero angular momentum 
component parallel to the magnetic field. In
this case (\ref{hartree2}) simplifies to the {\it Magnetic Hartree
density functional}
\beq\label{densfunc}
\Edens_\beta[\rho]=\int\left(|\nabla\sqrt\rho|^2+\frac{\beta^2}4
r^2\rho -\beta\rho-\frac 1{|\x|}\rho\right)d^3\x+D[\rho,\rho],
\eeq
where $r$ is the radial coordinate perpendicular to the magnetic
field. This functional is the restriction of (\ref{hartree2}) to
density matrices of the form $|\sqrt\rho\rangle\langle\sqrt\rho|$.
We will show in the next section that both $\E$ and $\Edens$ have
the same ground state energy and density, so one could
alternatively define $E^{\rm MH}$ as the infimum of $\Edens$.

The energy asymptotics for Region 2 are stated in the following
theorem:
\begin{thm}[Energy asymptotics for Region 2]\label{nrg2}
If $N\to \infty$  with $\lambda$ and $\beta$ fixed, then
\beq
\lim_{N\to\infty}\frac{1}{N}E(N,\lambda,\beta)=
\frac{1}{\lambda}E^{\rm MH}(\lambda,\beta).
\eeq
\end{thm}
\bigskip

In {\bf Region 3} the ground state of the atom is squeezed into a
needle with diameter - in the scaled coordinates - $\sim \beta
^{-1/2}$. The Coulomb interaction of the confined particles acts
along the needle effectively like a one dimensional delta
function, with coupling constant $\sim \ln \beta $. It thus
dictates the typical extension $\sim (\ln \beta )^{-1}$ of the
ground state wave function in the direction of the field, and the
typical energies as $\sim (\ln \beta )^{2}$. This effective
reduction to a one dimensional system has been discussed in
\cite{LSY94a, JY96, BSY00}; see also \cite{BRW99} for related
studies. In the appropriate scaling, $z\sim (\ln \beta )\xpar$,
with $\xpar$ the coordinate in the direction parallel to the
magnetic field, the theory is a Hartree theory for a one
dimensional model. It is identical to the theory for Region 5 of
fermionic electrons, which has been studied in \cite{LSY94a},
including an exact solution of the ground state problem. Its
energy functional has been called the {\it Hyper Strong}
Functional; it is
\beq\label{hartree3}
\mathcal{E}^{\rm HS}[\rho ]=\int (\frac{d}{dz} \rho
^{1/2}(z))^2dz- \rho (0) + \frac{1}{2}\int \rho (z)^2dz,
\eeq
with its ground state energy defined as
\beq\label{nrg3a}
E^{\rm HS}(\lambda)=\inf_{\rho, \hspace{4pt} \int \rho =
\lambda}\mathcal{E}^{\rm HS}[\rho ].
\eeq
We will prove
\begin{thm}[Energy asymptotics for Region 3]\label{nrg3}
If $N\to \infty$ and \newline $\beta = \beta (N)\to \infty$ with
$\lambda$ fixed, then
\beq
\lim_{N\to\infty}\frac{E(N,\lambda,\beta)}{N(\ln \beta)^2}=
\frac{1}{\lambda}E^{\rm HS}(\lambda).
\eeq
\end{thm}

The difference between bosonic atoms in our Region 3 to fermionic
atoms in the fermion-Region 5 is in the condition of applicability
of HS-theory. The Pauli principle demands one needle for each
electron, electrostatics makes them lying side by side. The
fermionic atom is thus a bundle of $N$ needles, with total
diameter - in unscaled coordinates and with unscaled parameters -
$\sim N^{1/2} B^{-1/2}$. The condition for validity of HS-Theory
is that the diameter of the atom, in the directions of $\xperp$,
perpendicular to the field, is much smaller than the
characteristic length of Coulombic quantum effects, $\sim Z^{-1}$.
This condition is therefore $B \gg N Z^2$ for fermionic atoms.
Bosonic electrons may all occupy the same needle. For them, the
particle number does therefore not appear in the condition, which
is now $B \gg Z^2$.

\bigskip

In the investigation of the limits we exploit four {\bf principles}:

\begin{enumerate}
\item[(i)] Restriction to independent particles,
\item[(ii)] Spatial concentration near the center,
\item[(iii)] Concentration in the lowest Landau band, and
\item[(iv)] High field limit of the Coulomb-interaction.
\end{enumerate}

The principle (i) is fundamental for the validity of the Hartree
theories: The atom with many interacting particles will be
compared to models with independent particles in an effective mean
field.
\newline
The spatial concentration (ii) had not to be stressed in systems
without magnetic fields. In the studies of fermionic electrons,
\cite{LSY94a, BSY00}, it has been proven as a consequence of the
superharmonicity of the repulsive interactions. We will also use
this superharmonicity, but in a different way: It implies the
vanishing of the parallel component of the angular momentum in the
state which minimizes the Hartree energy. Since one of the
consequences is the absence of an ``angular momentum barrier'', it
can also be viewed as a spatial concentration of the bound
electrons.
\newline
The principles (iii) and (iv) are effective in Region 3, in the
limit $\beta \to \infty$.

In the investigations of bosonic ``electrons'' we could probably
have mimicked the procedure of \cite{LSY94a}, with some changes
due to the Bose statistics. Our procedure relies in fact heavily
on the same methods, but we combine them in a new way: The
principles (ii), (iii) and (iv) mentioned above are studied for
single particles in effective mean fields. In the study of many
particle systems, we begin with the reduction to independent
particles. But this has to be done in a subtle way, anticipating
the limits which have to follow. We do this by extending the
method of \cite{BSY00}.

The Hartree theories will be discussed in Sect.
\ref{hartreetheory}, including the restriction to zero angular
momentum in Sect. \ref{min0}. The confinement to the lowest Landau
band, relevant for Region 3, is treated in Sect.
\ref{lowestlandauhartree} and the following subsection. In
Sect.~\ref{highnlimit} Hartree theory is proven to be the limit of
many particle quantum mechanics, as it is formulated in the
Theorems \ref{nrg1}, \ref{nrg2} and \ref{nrg3}. Subsection
\ref{restindependent} treats the restriction to the independent
particle model. Finally, some results on the states and on ``Bose
condensation'' are presented in Subsections \ref {convdens} and
\ref{bosecond}.

In the investigations of the limiting procedures, we are
interested in the physical dimensions of estimates and bounds, not
about numerics. We use ``$C$'' or ``$const.$'' for all the
numerical constants.

\section{Hartree theory}\label{hartreetheory}
\subsection{Definitions and basic properties} \label{hartreedef}
{\bf Definitions}: The Hartree functional without a magnetic
field, $\mathcal{E}^{\rm H}[\rho ]$ in (\ref{hartree1}), is
defined for non-negative densities $\rho(\x ) \in
\Ll^1(\R^3,d^3\x)$ with the restriction that every component of
$\nabla\rho^{1/2}(\x )$ is an element of $\Ll^2(\R^3,d^3\x)$.
Analogously, the functional for Region 3, $\mathcal{E}^{\rm
HS}[\rho ]$ in (\ref{hartree3}), is defined for non-negative
densities $\rho(z) \in \Ll^1(\R,dz)$ with the restriction
$d(\rho^{1/2}(z))/dz \in \Ll^2(\R ,dz)$. The functional for Region
2, $\mathcal{E}^{\rm MH}[\Gamma ]$ in (\ref{hartree2}), is defined
for density-matrices, non-negative trace class operators $\Gamma$
acting on $\Ll^2(\R^3,d^3\x)$, with the restriction of a finite
magnetic-kinetic energy:
\beq\label{mke}
\Tr[H_\beta\Gamma]<\infty
\eeq
The associated density $\rg(\x)$ can be defined in
$\Ll^1(\R^3,d^3\x)$ as a norm convergent sum of integrable
densities $\sum_k w_k \rho_k$, by diagonalizing the density matrix
as
\beq \label{gdia}
\Gamma = \sum_k w_k |\psi_k\rangle \langle \psi_k |,
\eeq
with normalized $\psi_k$, and $\rho_k (\x) = |\psi_k(\x)|^2$.

The conditions of finiteness of the kinetic energies imply the
finiteness of potential energies, in all three regions: the
attraction is bounded by the kinetic energy, because of the
boundedness of the Coulomb potential (delta function potential in
Region 3) relative to the operator of kinetic energy, which is
proven with the stability of the hydrogen atom. Moreover, the repulsion
is bounded by attraction, because for $\sigma(\x)$ and $\rho(\x )$
both non-negative elements of $\Ll^1(\R^3,d^3\x)$,
\beq
2D[\sigma,\rho] < \lambda \sup_{\y} A_{\y}[\rho],
\label{aboundr}
\eeq
with $\lambda = \int \sigma (\x)d^3\x $ and $A_{\y}[\rho] = \int
\frac{1}{|\x -\y|}\rho (\x)d^3\x $. The analogous inequality for the HS
theory is  
\beq
\int \sigma (z)\rho (z)dz < \lambda \sup_y \rho (y),
\eeq
for $\sigma(z)$ and
$\rho(z)$ both non-negative elements of $\Ll^1(\R,dz)$ and
$\lambda = \int \sigma (z)dz$.

\bigskip

In the {\bf variational principles} which define the Hartree
energies, the restrictions $\|\rho\|_1=\lambda$ and
$\Tr[\Gamma]=\lambda$ can be weakened to $\|\rho\|_1\leq\lambda$
and $\Tr[\Gamma]\leq\lambda$, because one can always \lq\lq move
some charge to infinity\rq\rq. This will be used for the proof of
the existence of a minimizer for the energy, and this in return
means that ``$\inf$'' can be replaced by \lq\lq $\min$\rq\rq.

In the following discussion we will explicitly study the magnetic
Hartree theory. All the results which do not refer to the
dependence on $\beta$ are also valid - in their essence of
physical meaning, with some changes in the mathematics - for the
standard and the HS theory. The proofs can be transfered, keeping
their structure, but changing the mathematical expressions. This
is an indication, that the physics behind the arguments is often
the same. We point out, in particular, that the delta potential
has the same scaling properties as the Coulomb potential.

Introducing more non-negative parameters, we will study the {\it
extended energy functional}
\beq\label{hartree4}
\E_{\lambda,\beta,\zeta ,\alpha}[\Gamma]=
\lambda\Tr[(H_\beta-\beta)\Gamma]-\lambda \int\frac{\zeta
}{|\x|}\rg(\x)d^3\x+\alpha\lambda^2D[\rg,\rg],
\eeq
and its ground state energy
\beq\label{nrg4a}
E^{\rm MH}_{\rm ext}(\lambda,\beta,\zeta ,\alpha)= \inf_{\Gamma ,
\hspace{4pt} \Tr[\Gamma] \leq 1}\E_{\lambda,\beta,\zeta
,\alpha}[\Gamma].
\eeq
By scaling one verifies that the energies are related through
\beq\label{rel}
E^{\rm MH}_{\rm ext}(\lambda,\beta,\zeta ,\alpha) =\frac{\zeta
^3}{\alpha}E^{\rm MH}(\frac{\alpha}{\zeta } \lambda,\frac{1}{\zeta
^2}\beta).
\eeq
{\bf Monotonicity, convexity and concavity properties}.

\begin{enumerate}
\item[(i)] Since the functional $\E_{\lambda,\beta,\zeta ,\alpha}[\Gamma]$
is decreasing in $\zeta $, increasing in $\alpha$ and jointly
linear in $(\zeta ,\alpha)$, the Hartree energy $E^{\rm MH}_{\rm
ext}$ is decreasing in $\zeta $, increasing in $\alpha$ and
jointly concave in $(\zeta ,\alpha)$.
\item[(ii)] Since the functional $\E_{\lambda,\beta,\zeta ,\alpha}[\Gamma]/\lambda$
is increasing in $\lambda$ and jointly linear in $(\zeta
,\lambda)$, the {\bf energy per unit charge} of the electron
cloud, $E^{\rm MH}_{\rm ext}/\lambda$, is increasing in $\lambda$
and jointly concave in $(\zeta ,\lambda)$.\newline Because of
(\ref{rel}), these properties are actually equivalent to (i).
\item[(iii)] Since moreover the functional $\E_{\beta}[\Gamma]$
is convex in $\Gamma$ (it is even strictly convex in $\rg$, since
$D[\rho,\rho]$ is strictly convex in $\rho$), and since the sets
$\{\Tr[\Gamma]\leq \lambda\}$ are ordered by inclusion, the
Hartree energy $E^{\rm MH}(\lambda,\beta)$ is decreasing and
convex in $\lambda$. For, if $\Gamma _n$ with $\Tr[\Gamma
_n]=\lambda$ and $\Upsilon _n$ with $\Tr[\Upsilon _n]=\bar
\lambda$ are minimizing sequences, it follows, that
$E^{\rm MH}((\lambda + \bar \lambda )/2,\beta)\leq %
\inf_n \E_{\beta}[(\Gamma _n + \Upsilon _n)/2]$.
\end{enumerate}
The convexity property justifies the definition of a {\bf critical
charge} $\lambda_c$, which a priori might be infinity, indicating
the maximal charge which the nucleus can bind: $E^{\rm MH}$ is
strictly decreasing for $\lambda \leq\lambda_c$, and constant for
$\lambda \geq\lambda_c$.

Moreover, the convexity of $E^{\rm MH}$ in $\lambda$ and the
concavity of $E^{\rm MH}/\lambda$ give upper and lower bounds on
$\partial ^2 E^{\rm MH}/\partial \lambda ^2$ , guaranteeing the
existence of the {\bf chemical potential}
\beq
\mu =\frac{\partial E^{\rm MH}}{\partial \lambda}
\eeq
as a continuous function of $\lambda$.

{\bf Comparison with the hydrogen atom:}
\newline
The extended functional (\ref{hartree4}) with $\alpha =0$ is
obviously related to the energy of the hydrogen atom described
with the Pauli Hamiltonian, where the magnetic moment of the
electron serves for the subtraction of $\beta$ in the ground state
energy:
\beq\label{hydro1}
E^{\rm MH}_{\rm ext}(\lambda,\beta,\zeta ,0)= \lambda E^{\rm
hyd}(\beta,\zeta ).
\eeq
We have the following bounds:

\begin{proposition}[Energies of hydrogen as bounds]\label{hydroprop}
For $\lambda >0$ and $\alpha >0$
\beqa
E^{\rm MH}_{\rm ext}(\lambda,\beta,\zeta ,\alpha)&>& \lambda
E^{\rm hyd}(\beta,\zeta ) \\ E^{\rm MH}_{\rm
ext}(\lambda,\beta,\zeta ,\alpha)&<&\lambda E^{\rm
hyd}(\beta,\zeta -\lambda \alpha /2). \label{hydupbd}
\eeqa
\end{proposition}
\begin{proof}
The first inequality follows from the strict positivity of
$D[\rho,\rho]$. The upper bound can be given by choosing the
projection onto the ground state of the Hamiltonian
$H_{\beta}-(\zeta -\lambda \alpha /2)/|\x|$, as a test density
matrix $\Gamma$. Then we apply the bound to the repulsion by
attraction (\ref{aboundr}), together with the observation, that
the nucleus has to be at the point of the maximum of the potential
$A_y[\rho]$. (Otherwise the energy could be lowered by shifting
$\rho$).
\end{proof}
We remark that the bound (\ref{hydupbd}) is of no use for $\lambda
\alpha \sim 2\zeta $ or larger. In this case one can use the
monotone decrease of the energy $E^{\rm MH}_{\rm ext}$ in
$\lambda$, and bound $E^{\rm MH}_{\rm ext}$ by $\min_{\lambda}
\{\lambda E^{\rm hyd}(\beta,\zeta -\lambda \alpha /2)\}$.
We will use these bounds in Subsection \ref{lowestlandauhartree}.

As an obvious consequence of these bounds, using also the
continuity of the hydrogen energy in $\zeta $, we add

\begin{rem}[The limit $\lambda \to 0$]\label{ll0}
In the limit $\lambda \to 0$ the energy per unit charge,
$E^{\rm MH}_{\rm ext}(\lambda,\beta,\zeta ,\alpha)/\lambda$,
converges  to the energy
of the hydrogen atom, $E^{\rm hyd}(\beta,\zeta )$.
\end{rem}

The bounds of Proposition \ref{hydroprop} will actually be used in
a simplified form:
\begin{lem}[Simple bounds]\label{simple}
\beqa
E^{\rm MH}(\lambda,\beta)&\geq&-(1/4+\beta)\lambda, \\ E^{\rm
MH}(\lambda,\beta)&\leq&-(1/4)\lambda (1-\lambda /2)^2,  \qquad
{\rm for} \quad \lambda \leq 2.
\eeqa
\end{lem}
\begin{proof}
Applying the diamagnetic inequality \cite{S76} to the lower bound
in proposition (\ref{hydroprop}), we get
\beq\label{214}
H_{\beta}-\beta-\frac{1}{|\x|} \geq\infspec
(H_0-\beta-\frac{1}{|\x|}) =-\frac{1}{4}-\beta.
\eeq
To get the upper bound, we apply Lieb's inequality
(Theorem A.1. in \cite{AHS78})
\beq \label{lieb}
\infspec (H_{\beta}-\beta +V) \leq \infspec (H_0 +V)
\eeq
to the upper bound in proposition (\ref{hydroprop}).
\end{proof}

\subsection{Minimizers}
\begin{thm}[Existence of a minimizer]\label{exist}
For each $\beta\geq 0$ and $\lambda>0$ there is a minimizer
$\Gamma^{\rm H}$ for $\E _\beta $ under the condition
$\Tr[\Gamma^{\rm H}]\leq\lambda $, i.e.
\beq
E^{\rm MH}(\lambda,\beta)=\E _\beta [\Gamma^{\rm H}].
\eeq
\end{thm}
\begin{proof}
We follow closely the proof of the analogous theorems 2.2 and 4.3
in \cite{LSY94a}. Let $\Gamma_n$ be a minimizing sequence for $\E
_\beta $ with $\Tr[\Gamma_n]\leq\lambda$. First note that
$\Tr[H_\beta\Gamma_n]$ is bounded above, because $\E _\beta
[\Gamma_n]$ is bounded above, and since the other contributions to
the energy are bounded relative to $H_\beta $. Now we show that
the magnetic-kinetic energy is bounded below by the 3-norm of
$\rho$: Using the diamagnetic inequality (the prerequisite for the
final result in \cite{S76})
\beq
\langle \psi ,(H_{\beta} +V)\psi \rangle \geq \langle |\psi |,(H_0
+V)|\psi |\rangle ,
\eeq
and the decomposition of $\Gamma$ as in (\ref{gdia}) ,
we get
\beq
\Tr[\nab ^2\Gamma]\geq
\sum_nw_n\int\left|\nabla| \psi _n|\right|^2.
\eeq
Moreover, using the Cauchy-Schwarz inequality,
\begin{eqnarray}\nonumber
|\nabla\rg|^2&=&\left|\sum_n2w_n| \psi _n|\nabla| \psi _n|\right|^2\\
&\leq&4\left(\sum_nw_n| \psi _n|^2\right)\left(\sum_nw_n
\left|\nabla| \psi _n|\right|^2\right).
\end{eqnarray}
Because $\nabla\rg=2\rg^{1/2}\nabla\rg^{1/2}$ this gives
\beq\label{l3}
\Tr[\nab^2\Gamma]\geq \int\left|\nabla\rg^{1/2}\right|^2\geq
3\left(\frac\pi 2\right)^{4/3}\left(\int\rg^3\right)^{1/3},
\eeq
where we have used the Sobolev inequality in the last step.

 We can then conclude that the
corresponding sequence $\rho_n\equiv\rho_{\Gamma_n}$ is bounded in
$L^3\cap\Ll^1$ and $\rho_n^{1/2}$ is bounded in $\Hh^1$.
Therefore, for each $p \in (1, 3]$, there exists a subsequence,
again denoted by $\rho_n$, that converges to some $\rho_\infty$
weakly in $\Ll^3\cap\Ll^p$, and pointwise almost everywhere. It
follows from weak convergence that $\rho_\infty\geq 0$ and
$\int\rho_\infty\leq\lambda$. From Fatou's lemma we infer that
\beq \label{dconv}
\liminf_{n\to\infty}D[\rho_n,\rho_n]\geq
D[\rho_\infty,\rho_\infty].
\eeq
Moreover, since $|\x|^{-1}\in\Ll^{3/2}+\Ll^{3+\eps}$ for every
$\eps>0$, and choosing $p$ the dual of $3+\eps$, we see that
\beq \label{aconv}
\lim_{n\to\infty}\int\frac{1}{|\x|}\rho_n
=\int\frac{1}{|\x|}\rho_\infty.
\eeq

Since the $\Gamma_n$'s are trace class operators on $\Ll^2$, we
can pass to a subsequence such that for some $\Gamma^{\rm H}$
\beq
\lim_{n\to\infty}\Tr[\Gamma_n A]=\Tr[\Gamma^{\rm H} A]
\eeq
for every compact operator $A$. (Here we used the Banach-Alaoglu
Theorem and the fact that the dual of the compact operators is the
trace class operators). In particular, we have
\beq\label{weak}
\Gamma_n\rightharpoonup\Gamma^{\rm H}
\eeq
in the weak operator sense. It is clear that $\Gamma^{\rm H}\geq 0$. Now
let $\phi_j\in \Cc_0^\infty(\R^3)$ be an orthonormal basis for
$\Ll^2$. Again by Fatou's Lemma
\beq
\Tr[\Gamma^{\rm H}]=\sum_j\langle\phi_j|\Gamma^{\rm H}|\phi_j\rangle\leq
\liminf_{n\to\infty}\Tr[\Gamma_n]\leq\lambda.
\eeq
In the same way one shows that
\begin{eqnarray}\nonumber
\Tr[\left(H_\beta-\beta\right)\Gamma^{\rm H}]&=&\sum_j\langle
\left(H_\beta-\beta\right)^{1/2}
\phi_j|\Gamma^{\rm H}|\left(H_\beta-\beta\right)^{1/2}\phi_j\rangle\\
&\leq&
\liminf_{n\to\infty}\Tr[\left(H_\beta-\beta\right)\Gamma_n].
\end{eqnarray}
It remains to show that $\rho_{\Gamma^{\rm H}}=\rho_\infty$. We already
mentioned that for some constant $C$
\beq\label{bound}
\Tr[H_\beta\Gamma_n]<C
\eeq
for all $n$. It follows from (\ref{weak}) that
\beq\label{weakly}
(1+H_\beta)^{1/2}\Gamma_n(1+H_\beta)^{1/2}\rightharpoonup
(1+H_\beta)^{1/2}\Gamma^{\rm H}(1+H_\beta)^{1/2}
\eeq
weakly on the dense set of $\Cc_0^\infty$ functions. Since the
operators are bounded by (\ref{bound}), (\ref{weakly}) holds
weakly in $\Ll^2$.

Now consider some $f\in\Cc_0^\infty$ acting as a multiplication
operator on $\Ll^2$. It is easy to see that $f$ is relatively
compact with respect to $-\Delta$, i. e. $f(1-\Delta)^{-1}$ is
compact. In fact, it is Hilbert-Schmidt, because the trace of its
square is given by
\beq
\int f(\x)f(\y)Y(\x-\y)^2d^3\x\, d^3\y
\eeq
with the Yukawa-Potential $Y(\x)=(4\pi|\x|)^{-1}\exp(-|\x|)$, and
this is bounded by Young's inequality. From \cite{AHS78}, Thm. 2.6,
we infer that
\beq
g=(1+H_\beta)^{-1/2}f(1+H_\beta)^{-1/2}
\eeq
is compact (it is even Hilbert-Schmidt). Thus there exists a
sequence $g_i$ of finite-rank operators which approximates $g$ in
norm. We have
\begin{eqnarray}\nonumber
\left|\int(\rho_n-\rh)f\right|&=&
\left|\Tr[(\Gamma_n-\Gamma^{\rm H})f]\right|\\ \nonumber
&\leq&\left|\Tr[(1+H_\beta)^{1/2}(\Gamma_n-\Gamma^{\rm H})
(1+H_\beta)^{1/2}g_i]\right|\\ & &+2(C+1)\lambda\|g-g_i\|,
\end{eqnarray}
where we have used (\ref{bound}). Hence $\rho_n\to\rh$ in the
sense of distributions. Because we already know that $\rho_n$
converges to $\rho_\infty$ pointwise almost everywhere, we
conclude that $\rh=\rho_\infty$. We have thus shown that there
exists a $\Gamma^{\rm H}$ with $\Tr[\Gamma^{\rm H}]\leq\lambda$ and
$\E_{\beta}[\Gamma^{\rm H}]\leq\liminf_{n\to\infty}\E_{\beta}[\Gamma_n]$,
from which we conclude that $\E_{\beta}[\Gamma^{\rm H}]=E^{\rm
MH}(\lambda,\beta)$.
\end{proof}

\begin{rem}
If $\Gamma^{\rm H}$ is unique, and given any minimizing sequence for $\E
_\beta $, the whole sequence converges weakly to $\Gamma^{\rm H}$.
\end{rem}

Although we cannot make an assertion about the uniqueness of
$\Gamma^{\rm H}$ yet, we can state the
\begin{proposition}[Uniqueness of the density]
The density $\rho^{\rm H}$ corresponding to the minimizer is unique.
\end{proposition}
\begin{proof}
This follows immediately from the strict convexity of $D[\rho,\rho]$.
\end{proof}

For $\lambda \leq \lambda _c$, the energy $E^{\rm MH}$ is a
strictly decreasing function of $\lambda$. So the minimizers for
different $\lambda \leq \lambda _c$ are different, and they have
the normalization $\Tr[\Gamma]=\lambda$. No part of the electron
cloud has to be moved to infinity. The strict convexity of
$D[\rho,\rho]$ implies now, for $\lambda \leq \lambda _c$, a
strict convexity of $E^{\rm MH}$ as a function of $\lambda$.

\subsection{Some physical quantities and their interrelations}
Since the densities of the minimizers are unique,
the contributions to the Hartree energy, $\int \rho /|\x|$ and
$D[\rho,\rho]$, are fixed.
We denote them as attraction $A$ and repulsion $R$,
suppressing the dependence on the parameters.
As a consequence, also $K$, the kinetic-magnetic energy, is fixed.
Inserting a minimizer into (\ref{hartree2}) we get
\beq
E=K-A+R.
\eeq

To deduce an analogue to the Feynman-Hellman theorem, we observe
the following inequality: Consider two different parameters,
$\zeta $ and $\bar \zeta $, with corresponding minimizers $\Gamma$
and $\bar \Gamma$, and their densities $\rho$ and $\bar \rho$. All
the other parameters are fixed. Insert $\bar \Gamma$ into the
functional (\ref{hartree4}) with parameter $\zeta $, to conclude
\begin{eqnarray} \nonumber
E^{\rm MH}_{\rm ext}(\lambda,\beta,\zeta ,\alpha)&<&
\E_{\lambda,\beta,\zeta ,\alpha}[\bar \Gamma] \\ &=&
\E_{\lambda,\beta,\bar \zeta ,\alpha}[\bar \Gamma]- (\zeta -\bar
\zeta )\int\frac{1} {|\x|}\bar \rh  \nonumber \\ &=& E^{\rm
MH}_{\rm ext}(\lambda,\beta,\bar \zeta ,\alpha)-(\zeta -\bar \zeta
)\int\frac{1}{|\x|}\bar \rho.
\end{eqnarray}
Together with the same argument, where $\zeta $ and $\bar \zeta $
are exchanged, we get, for $\zeta >\bar \zeta $,
\beq
-\int\frac{1}{|\x|}\rho <
\frac{E^{\rm MH}_{\rm ext}(\lambda,\beta,\zeta ,\alpha)-%
E^{\rm MH}_{\rm ext}(\lambda,\beta,\bar \zeta ,\alpha)}{\zeta
-\bar \zeta }< -\int\frac{1}{|\x|}\bar \rho.
\eeq
We conclude, using the continuity (\ref{aconv}), where the
$\rho_n$ are now the unique densities of the minimizers for $\bar
\zeta _n \to \zeta $, that the Hartree energy is differentiable in
$\zeta $, and
\beq
\left.\frac{\partial E^{\rm MH}_{\rm ext}}{\partial \zeta }
\right|_{\zeta =1}=-A.
\eeq
In the same way, using the inequality (\ref{dconv}) instead of
(\ref{aconv}), we show that
\beq
\left.\frac{\partial E^{\rm MH}_{\rm ext}}{\partial \alpha}
\right|_{\alpha =1}=R.
\eeq
Having established the differentiability in $\lambda$, $\zeta $
and in $\alpha$, the scaling relation (\ref{rel}) implies
differentiability in $\beta$. So, the {\bf magnetic moment $\theta
$},
\beq
\theta =\frac{\partial E^{\rm MH}}{\partial \beta},
\eeq
is well defined as a function of $\beta$.

Taking the partial derivatives of (\ref{rel}) in $\alpha$ and $\zeta $
at $\alpha =1$ and $\zeta =1$ gives
\beqa
R&=&-E+\lambda \mu , \label{mu}\\
-A&=&3E-\lambda \mu -2\beta \theta . \label{mutheta}
\eeqa
The relation (\ref{mu}) is, because of the negativity of the
chemical potential $\mu$ for $\lambda < \lambda_c$ , a {\bf virial
inequality},
\beq \label{virin}
R<|E|,
\eeq
turning into an equality at $\lambda = \lambda_c$.
Inserting (\ref{mu}) into (\ref{mutheta}), we get
\beq \label{theta}
\beta \theta=\frac{K-|E|}{2}.
\eeq
The equation (\ref{mu}) and the inequality (\ref{virin}) are also
valid in the Regions 1 and 3, where there is a scaling relation
analogous to (\ref{rel}). But this relation is without a $\beta$,
so instead of (\ref{theta}), in Regions 1 and 3 the well known
{\bf virial equality}
\beq \label{virial}
K=|E|=\frac{A-R}{2}
\eeq
holds (see also \cite{LSY94a, B84}). Moreover, for $\lambda
=\lambda _c$ we have $|E|:K:A:R=1:1:3:1$.

\subsection{The linearized theory}
Since the density $\rho^{\rm H}$ of a minimizer is unique and
integrable, we can define a linearized Hartree functional by
\beq
\tE[\Gamma]=\Tr[H^{\rm H}\Gamma],
\eeq
with the one-particle Hartree Hamiltonian
\beq\label{hartop}
H^{\rm H} \equiv H_\beta-\beta-\Phi^{\rm H}(\x).
\eeq
Here the Hartree potential $\Phi^{\rm H}(\x)$ is given by
\beq
\Phi^{\rm H}(\x)=\frac{1}{|\x|}-\rho^{\rm H}*\frac{1}{|\x|}.
\eeq
Note that $H^{\rm H}$ depends on $\lambda$ only via $\rho^{\rm H}$.
\newline
Since $\Phi^{\rm H}$ is in $\Ll^2+\Ll^\infty_\eps$, it is relative
compact with respect to $H_\beta$ \cite{AHS78}, so we know that
the essential spectrum of $H^{\rm H}$ is $[0,\infty)$.

\begin{lem}[Linear Hartree functional]\label{linear}
Let $\Gamma^{\rm H}$ be a minimizer of $\E_{\beta}$ under the constraint
$\Tr[\Gamma]\leq\lambda$. Then $\Gamma^{\rm H}$ also minimizes $\tE$
(under the same constraint).
\end{lem}
\begin{proof}
(We proceed essentially as in \cite{LSY94a}). For any $\Gamma$
\beq
\E_{\beta}[\Gamma]=\tE[\Gamma]+D[\rg-\rho^{\rm H},\rg-\rho^{\rm H}]-D[\rho^{\rm H},\rho^{\rm H}].
\eeq
Especially, for all $\delta>0$,
\begin{eqnarray}\nonumber
\E_{\beta}[(1-\delta)\Gamma^{\rm H}+\delta\Gamma]&=&
(1-\delta)\tE[\Gamma^{\rm H}]+\delta\tE [\Gamma]-D[\rho^{\rm H},\rho^{\rm H}]\\
&&+\delta^2D[\rg-\rho^{\rm H},\rg-\rho^{\rm H}].
\end{eqnarray}
Now if there exists a $\Gamma_0$ with $\Tr[\Gamma_0]\leq\lambda$
and $\tE[\Gamma_0]<\tE[\Gamma^{\rm H}]$ we can choose $\delta$ small
enough to conclude that
\beq
\E_{\beta}[(1-\delta)\Gamma^{\rm H}+\delta\Gamma_0]<\tE
[\Gamma^{\rm H}]-D[\rho^{\rm H},\rho^{\rm H}]=\E_{\beta}[\Gamma^{\rm H}],
\eeq
which contradicts the fact that $\Gamma^{\rm H}$ minimizes $\E_{\beta}$.
\end{proof}
The {\bf Hartree equation}, the Euler-Lagrange equation
corresponding to the minimization of the functional $\E_{\beta}$,
is
\beq
H^{\rm H} \Gamma = \mu \Gamma .
\eeq
The ground state energy of $H^{\rm H}$ is given by the chemical
potential $\mu (\lambda ,\beta )=(E+R)/\lambda$. For the
overcritical values of $\lambda \geq \lambda_c$ the ground state
energy $\mu$ of $H^{\rm H}$ is $0$, and there is just one density
corresponding to the minimizer.

We can now ensure that $\lambda_c=\lambda_c(\beta)$ is not too
small. In fact we state

\begin{lem}[Lower bound on the critical charge]\label{critlow}
\beq
\lambda_c(\beta)>1\qquad{\rm for\ all\ \beta\geq 0}.
\eeq
\end{lem}
\begin{proof}
For $\beta=0$ this was shown in \cite{BL83}, and $\lambda_c(0)$ was
computed numerically to be $1.21$ in \cite{B84}.

Fix $\beta>0$. We assume that $\lambda_{\rho}\equiv
\int\rho^{\rm H}\leq 1$, and will show that
$H^{\rm H}$ has an eigenvalue strictly below zero. Using $\psi(r,z)=\exp
(-\beta r^2/4-a|z|)$ with some $a>0$ as a
(not normalized) test function we compute
\begin{eqnarray}\nonumber
\langle\psi|H^{\rm H}\psi\rangle&=&\frac{2\pi}\beta  a-2\pi\int\Phi^{\rm H}(r,z)
e^{-\half \beta r^2-2a|z|}rdrdz\\
&\leq&\frac{2\pi}\beta  a-\int\rho^{\rm H}(\x)
\left(\phi({\bf 0})-\phi(\x)\right)d^3\x,
\end{eqnarray}
where $\phi(\x)$ is the potential generated by the charge
distribution $\exp(-\half \beta r^2-2a|z|)$, i.e.
\beq
\phi(\y)=\int\frac{e^{-\half \beta r^2-2a|z|}}{|\x-\y|}d^3\x.
\eeq
Note that $\phi (\0) >\phi (\y)$ for $\y\not= \0$, so we can
choose $a$ small enough to conclude that $H^{\rm H}$ has an eigenvalue
strictly below zero and binds more charge than $\lambda_{\rho}$.
\end{proof}

Lemma \ref{critlow} shows that in Hartree theory there are always
negative ions. The following lemma gives an upper bound on the
maximal negative ionization.

\begin{lem}[Upper bound on the critical charge]\label{critup}
For some con\-stant $C$ independent of $\beta$
\beq
\lambda_c(\beta)\leq 2+\frac 12\min\left\{1+\beta,
C\left(1+[\ln\beta]^2\right)\right\}.
\eeq
\end{lem}
\begin{proof}
This follows from Lemma \ref{critlem} and the bound on $N_c$
given in \cite{S00} (see Remark \ref{critrem} below).
\end{proof}

\subsection{The minimizer has $\Lpar =0$} \label{min0}
Let $H$ be the  operator
\beq\label{defhr}
H=H_\beta-\beta-\Phi(\x),
\eeq
where the  potential is given by
\beq
\Phi(\x)=\frac 1{|\x|}-\frac 1{|\x|}*\rho.
\eeq
The function $\rho$ is assumed to be {\it axially symmetric},
non-negative, and $\rho\in\Ll^1\cap\Ll^q$, where $q>3/2$. This
implies that $|\x|^{-1}*\rho$ is a bounded, continuous function
going to zero at infinity. The operator $H$ is essentially
self-adjoint on $\Cc_0^\infty(\R^3)$. Since
$\Phi\in\Ll^2+\Ll_\eps^\infty$, it is a relatively compact
perturbation of $H_\beta$ \cite{AHS78}, so we know that the
essential spectrum of $H$ is given by $[0,\infty)$. The symmetry
of $\rho$ implies that $H$ is axially symmetric, i.e. it commutes
with the rotations generated by the parallel component of the
angular momentum,
\beq
\Lpar=-i\left(x\dy -y\dx\right).
\eeq

For $m\geq 0$ let now $\Psi_m$ be a ground state of $H$, if there
is one, with angular momentum $\Lpar\Psi_m=-m\Psi_m$. Note that we
can restrict ourselves to considering non-negative $m$'s, since
$(H \upharpoonright \Lpar=-m)$ is antiunitarily equivalent to
$(H+2\beta m \upharpoonright \Lpar=m)$  by complex conjugation, so
in the ground state $m$ is certainly non-negative. Writing
\beq\label{polar}
\Psi_m(\x)=e^{-im\varphi}r^m f(r,z),
\eeq
where $(r,\varphi)$ denote polar coordinates for $(x,y)$, we see
that $f$ is a ground state for
\beq
\widetilde H_m=-\frac{\partial^2}{\partial r^2}-\frac{2m+1}r
\frac{\partial}{\partial r}-\frac{\partial^2}{\partial
z^2}+\frac{\beta^2}4 r^2-(m+1)\beta-\Phi(r,z)
\eeq
on $\Ll^2(\R^3,r^{2m}d\x)$. If $\widetilde H_m$ has a ground
state, it is unique and strictly positive (\cite{RS78}; to apply
the theorems therein note that the first three terms in
$\widetilde H_m$ are just the radial part of the Laplacian in
$2m+3$ dimensions). So $\Psi_m$ is the only ground state of $H$
with $\Lpar=-m$. Moreover, $f$ is a bounded, continuous function
\cite{LL97}, and hence
\beq \label{smallr}
|\Psi_m (r,\phi ,z)|\leq {\rm const.}r^m,
\eeq
for small $r$ and some constant independent of $z$.
In particular, $\Psi_m({\bf 0})=0$ if $m>0$.

We now are able to prove the main result of this subsection:

\begin{thm}[$m=0$ in the ground state of $H$]\label{m0}
Let $H$ be a Hamiltonian as in (\ref{defhr}), with $\rho$ axially
symmetric and non-negative. If \ $\infspec H$ is an eigenvalue,
the corresponding eigenvector is unique and has zero angular
momentum, for all values of $\beta\geq 0$.
\end{thm}
\begin{proof}
For $\rho=0$, i.e. the hydrogen atom, this was shown in \cite{AHS81} (and
also in \cite{GS95}). So
we can restrict ourselves to considering the case $\int \rho >0$.

Let $\Psi_m$ be a normalized ground state for $H$, with angular
momentum $-m$. Define
\beq
f(\B)=-\int |\Psi_m(\x)|^2 \Phi(\x-\B)d^3\x.
\eeq
The function $f$ is continuous and bounded. It achieves its
minimal value at $\B=\0$, because otherwise one could lower the
energy by translating $|\Psi_m|^2$. Strictly speaking,
\beq
f(\B)-f(\0)=\langle\widetilde\Psi_m|H|\widetilde\Psi_m\rangle
-\langle\Psi_m|H|\Psi_m\rangle\geq 0,
\eeq
with $\B=(b_1,b_2,b_3)$ and
\beq\label{phase}
\widetilde\Psi_m(\x)=e^{-\mbox{$\frac i2$} \beta (b_2 x-b_1
y)}\Psi_m(\x+\B).
\eeq
The phase in (\ref{phase}) is chosen such that the kinetic energy
remains invariant. (Note that translating $H_\beta$ is equivalent
to changing the gauge of the magnetic potential.)

In the sense of distributions,
\beq\label{dist}
\Delta f(\B)=4\pi\left(|\Psi_m(\B)|^2-|\Psi_m|^2*\rho_-(\B)\right),
\eeq
with $\rho_-(\x)\equiv\rho(-\x)$.
The function $|\Psi_m|^2*\rho_-$ is pointwise strictly positive and
continuous.
Assume now
that $m>0$. Since, by (\ref{smallr}), $|\Psi_m(\x)|\leq C|\x|^m$ for some
$C>0$, there is an $R>0$
such that
\beq
\Delta f(\B)<0 \quad {\rm for}\quad |\B|< R,
\eeq
i.e. $f$ is superharmonic in some open region containing $\0$.
This contradicts the fact that $f$ achieves its minimum at
$\B=\0$. As a consequence, a ground state of $H$ must have angular
momentum $m=0$.
\end{proof}

An analogous result holds also for the restriction of $H$ to the
lowest Landau band. This fact will be used in the proof of Theorem
\ref{nrglarge}.

\begin{cl}[$m=0$ in the ground state of $\Pi_0 H \Pi_0$]\label{conf0}
Let $\Pi_0$ be \newline the projector onto the lowest Landau
band. If\, $\Pi_0H\Pi_0$
has a ground state, the ground state wave function $\Psi_0$ is given by
\beq\label{psi0}
\Psi_0(\x)=\sqrt\frac\beta{2\pi}\exp\left(-\frac\beta 4r^2\right)\psi(z)
\eeq
for some $\psi(z)$ with $\int|\psi(z)|^2dz=\|\Psi_0\|_2^2$.
\end{cl}
\begin{proof}
Mimicking the proof of the last theorem, we see that if\,
$\Pi_0H\Pi_0$ has a ground state, the corresponding wave function
$\Psi_0=\Pi_0\Psi_0$ has $\Lpar=0$. Hence it is given by
(\ref{psi0}).
\end{proof}

Let now $\rho$ be the Hartree density $\rho^{\rm H}$, the unique density
of the minimizer of $\E_\beta $ (depending on $\lambda$ and
$\beta$). Applying Lemma (\ref{linear}) and the theorem above, we
immediately get

\begin{cl}[Uniqueness of $\Gamma^{\rm H}$]\label{unique}
The functional $\E _\beta $ has a unique minimizer $\Gamma^{\rm H}$
under the condition $\Tr[\Gamma]\leq \lambda$, which is
proportional to the projection onto the positive function
$\sqrt{\rho^{\rm H}}$. In particular, $\Gamma^{\rm H}$ has rank $1$. Moreover,
$\rho^{\rm H}$ minimizes the is density functional (\ref{densfunc})
under the condition $\int\rho\leq\lambda$. It is $\Cc^\infty$ away
from the origin, continuous at $\x=\0$, and strictly positive.
\end{cl}

The properties of $\rho^{\rm H}$ stated above follow from the fact that
$\rho^{\rm H}$ minimizes $\Edens_\beta$, and therefore satisfies the
variational equation
\beq\label{varrho}
-\Delta\sqrt{\rho^{\rm H}}+\left(\frac{\beta^2}4 r^2-
\beta-\Phi^{\rm H}\right)\sqrt{ \rho^{\rm H}}=\mu\sqrt{\rho^{\rm H}},
\eeq
where $\mu$ is the chemical potential of the MH theory. Note that
Corollary \ref{unique} proves the assertion made in the
Introduction that both $\E$ and $\Edens$ have the same ground
state energy and density.

\section{The limits of Region 2} \label{betalimits}

\subsection{The limit of very weak magnetic fields} \label{lowfieldlimit}

The limit $\beta \to 0$ is rather easy to handle:

\begin{thm}[Hartree energy for small $\beta$]\label{nrgsmall}
In the limit $\beta \to 0$,
\beq\label{ob}
E^{\rm MH}(\lambda,\beta)=E^{\rm
H}(\lambda)-\beta\lambda+O(\beta ^2).
\eeq
\end{thm}

\begin{proof}
Fix $\lambda>0$, and let $\rho_\beta$ be the minimizer of the
density functional $\Edens_\beta$ under the constraint
$\int\rho\leq\lambda$. Using $\rho_0$ as a trial function we
get
\beq
E^{\rm MH}(\lambda,\beta)\leq \Edens_\beta[\rho_0]=E^{\rm
MH}(\lambda,0)-\beta\lambda+\frac{\beta^2}4\int r^2\rho_0.
\eeq
It is well known that $\rho_0$ falls off more quickly that the
inverse of any polynomial, so $\int r^2\rho_0$ is finite
\cite{vw2}.

For the converse, we estimate
\beq\label{converse}
E^{\rm MH}(\lambda,\beta)=\Edens_\beta[\rho _\beta] \geq
\Edens_0[\rho _\beta]-\beta\lambda \geq E^{\rm
MH}(\lambda,0)-\beta\lambda.
\eeq
The observation $E^{\rm MH}(\lambda,0)=E^{\rm H}(\lambda)$ then
leads to (\ref{ob}).
\end{proof}

\subsection{Lowest Landau band confinement in Hartree
theory}\label{lowestlandauhartree}

We now show that for large $\beta$ most of the charge is confined to the lowest Landau
band. To do this, we write the Hartree functional as
\beq
\E_\beta[\Gamma]=\Tr\left[\left(H_\beta-\beta-\frac
1{|\x|}\right)\Gamma\right]
+\half\Tr_2\left[\Gamma\otimes\Gamma \frac 1{|\x-\y|}\right],
\eeq
where $\Tr_2$ means the trace over the doubled space
$\Ll^2(\R^3,d^3\x)\otimes \Ll^2(\R^3,d^3\y)$. Let $\Pi_0$ be the
projector onto the lowest Landau band, and let $\Pi_>=1-\Pi_0$. We
will use the decomposition
\beq\label{35}
H_\beta-\beta=\Pi_0(H_\beta-\beta)\Pi_0+\Pi_>(H_\beta-\beta)\Pi_>
\eeq
and
\beq
\frac 1{|\x|}=\Pi_0\frac 1{|\x|}\Pi_0+\Pi_>\frac 1{|\x|}\Pi_>+
\Pi_0\frac 1{|\x|}\Pi_>+\Pi_>\frac 1{|\x|}\Pi_0.
\eeq
The off-diagonal terms can be bound using
\beq
\left(\sqrt\eps\Pi_0-\frac 1{\sqrt\eps}\Pi_>\right)\frac 1{|\x|}
\left(\sqrt\eps\Pi_0-\frac 1{\sqrt\eps}\Pi_>\right)\geq 0
\eeq
for some $0<\eps<1$, with the result that
\beq\label{38}
\frac 1{|\x|}\leq(1+\eps)\Pi_0\frac 1{|\x|}\Pi_0+\left(1+\frac
1\eps\right)
\Pi_>\frac 1{|\x|}\Pi_>.
\eeq
In the same way one shows that, see \cite{LSY94a},
\begin{eqnarray}\nonumber
\frac 1{|\x-\y|}&\geq& (1-3\eps)\Pi_0\otimes\Pi_0\frac
1{|\x-\y|}\Pi_0\otimes\Pi_0\\ \nonumber
&+&\left(1-\frac 3\eps\right)\Pi_>\otimes\Pi_>\frac
1{|\x-\y|}\Pi_>\otimes\Pi_>\\ \nonumber
&+&\left(1-2\eps-\frac 1\eps\right)\Pi_0\otimes\Pi_>\frac
1{|\x-\y|}\Pi_0\otimes\Pi_>\\
&+&\left(1-\eps-\frac 2\eps\right)\Pi_>\otimes\Pi_0\frac
1{|\x-\y|}\Pi_>\otimes\Pi_0.
\end{eqnarray}
Moreover, we will use
\beq
\Tr_2\left[\Pi_{\#}\otimes\Pi_>\frac 1{|\x-\y|}\Pi_\#\otimes\Pi_>
\Gamma\otimes\Gamma\right]
\leq\Tr[\Pi_\#\Gamma]\sup_{\abf}\Tr\left[\Pi_>\Gamma\Pi_>\frac
1{|\x-\abf|}\right],
\eeq
where $\#$ stands for either $0$ or $>$. If we restrict ourselves to
considering $\Gamma$'s
with $\Tr[\Gamma]\leq\lambda$ we therefore have
\begin{eqnarray}\nonumber
\E_\beta[\Gamma]&\geq&
(1-3\eps)\E_\beta[\Pi_0\Gamma\Pi_0]+3\eps\lambda\left(\frac
83\right)^2 \infspec\left(H_\beta-\beta-\frac 1{2|\x|}\right)\\
\label{abc} &+&\Tr[\Pi_>\Gamma]\left(\frac \beta
2+\inf_\abf\infspec\left(\half H_\beta- \frac 2\eps\frac
1{|\x|}-\frac{2\lambda}{\eps}\frac 1{|\x-\abf|}\right) \right),
\end{eqnarray}
where we have used that
\beq\label{311a}
\Pi_>(H_\beta-\beta)\Pi_>\geq\half\Pi_>(H_\beta+\beta)\Pi_>
\eeq
and that
\beq\label{39}
\Pi_0\left(H_\beta-\beta-\frac\delta {2|\x|}\right)\Pi_0\geq
\delta^2\Pi_0\left(H_\beta-\beta-\frac1 {2|\x|}\right)\Pi_0
\eeq
for $\delta\geq 1$, which can easily be seen by scaling
$z\to\delta^{-1}z$.

Now we use the comparison with the hydrogen atom, Proposition
(\ref{hydroprop}):
\beq\label{310}
\infspec\left(H_\beta-\beta-\frac 1{2|\x|}\right)\geq
\max\{1,\lambda^{-1}\}
E^{\rm MH}(\lambda,\beta).
\eeq
By the same argument as in the proof of this inequality,
we can set $\abf={\bf 0}$ in
(\ref{abc}). Using the diamagnetic inequality we see that $\half
H_\beta-c|\x|^{-1}\geq -\half c^2$, so finally
\begin{eqnarray}\nonumber
\E_\beta[\Gamma]&\geq&
(1-3\eps)\E_\beta[\Pi_0\Gamma\Pi_0]+3\eps\left(\frac 83\right)^2
\max\{1,\lambda\}E^{\rm MH}(\lambda,\beta)\\ \label{314} &+&\frac
12\left(\beta -
\frac{4(1+\lambda)^2}{\eps^2}\right)\Tr[\Pi_>\Gamma].
\end{eqnarray}
Now if $\beta$ is large enough, we can set
$\eps^2=4(1+\lambda)^2/\beta$ to conclude

\begin{lem}[Comparison with the confined Hartree theory]\label{compconf}
\beq\label{confconst}
E_{\rm conf}^{\rm MH}(\lambda,\beta)\leq E^{\rm
MH}(\lambda,\beta)(1-{\rm const.}(1+\lambda)^2\beta^{-1/2}),
\eeq
where
\beq
E_{\rm conf}^{\rm
MH}(\lambda,\beta)=\inf_{\Tr[\Pi_0\Gamma]\leq\lambda}\E_\beta[\Pi_0\Gamma
\Pi_0].
\eeq
\end{lem}

Note that the constant in (\ref{confconst}) can be chosen such that (\ref{confconst}) is
valid for all values of $\beta>0$ and $\lambda>0$.

\begin{rem} Equation (\ref{314}) can be used to estimate the part of the Hartree state
$\Gamma^{\rm H}$ not confined to the lowest Landau band. In fact,
if $\eps^2>4(1+\lambda)^2/\beta$ in (\ref{314}), and with $\Gamma=\Gamma^{\rm H}$, we get
\beq
\Tr[\Pi_>\Gamma^{\rm H}]\leq -{\rm const.}(1+\lambda)\frac {\eps E^{\rm
MH}(\lambda,\beta)} {\beta -4(1+\lambda)^2\eps^{-2}}.
\eeq
Using the simple lower bound on $E^{\rm MH}$, given in Lemma
\ref{simple}, and optimizing over $\eps$ gives
\beq
\Tr[\Pi_>\Gamma^{\rm H}]\leq {\rm const.}(1+\lambda)^3\beta^{-1/2}.
\eeq
\end{rem}

\subsection{Confinement for the mean field Hamiltonian}

An analogous result as in the previous subsection holds also for
the linearized theory. The following estimate will be used in
Subsection \ref{reg3}:

\begin{lem}[Confinement for the mean field
Hamiltonian]\label{confmean} Let $H$ be given as in (\ref{defhr}),
with $\rho\geq 0$ and $\int\rho\leq\lambda$. Then for all
$\beta>0$
\beq\label{320}
\infspec H \geq \infspec \Pi_0 H \Pi_0 +(2+\lambda)\beta^{-1/2}
E^{\rm hyd}(\beta,2).
\eeq
\end{lem}

\begin{proof}
For $0<\eps<1$ we use again (\ref{35}) and (\ref{38}), and the analogous
inequality in the other direction for $|\x|^{-1}*\rho$, to
conclude that
\begin{eqnarray}\nonumber
H &\geq& \Pi_0\left(H_\beta-\beta-(1+\eps)\frac 1{|\x|}+(1-\eps)
\frac 1{|\x|}*\rho
\right)\Pi_0 \\ \nonumber & & + \Pi_>\left(H_\beta-\beta-
(1+\eps^{-1})\frac 1{|\x|}+(1-\eps^{-1})
\frac 1{|\x|}*\rho\right)\Pi_>\\ \nonumber
&\geq& (1-\eps)\Pi_0 H \Pi_0+\eps \Pi_0\left(H_\beta-\beta-
\frac 2{|\x|}\right)\Pi_0
\\ & & + \half \Pi_>\left(H_\beta+\beta-\frac {4\eps^{-1}}{|\x|}-
\frac {2\eps^{-1}}{|\x|}*\rho\right)\Pi_>,
\end{eqnarray}
where we have also used (\ref{311a}).
With the aid of the inequality (\ref{aboundr}) one easily sees that
\beq
\infspec \left(H_\beta-\frac {4\eps^{-1}}{|\x|}- \frac
{2\eps^{-1}}{|\x|}*\rho\right) \geq \infspec \left(H_\beta-\frac
{(4+2\lambda)\eps^{-1}}{|\x|}\right).
\eeq
Using again $\half
H_\beta-c|\x|^{-1}\geq -\half c^2$ we finally get
\beq
H \geq (1-\eps) \Pi_0 H \Pi_0 + \eps
E^{\rm hyd}(\beta,2)+\half\Pi_>\left(\beta-(2+\lambda)^2
\eps^{-2}\right).
\eeq
In particular, if we choose $\eps=(2+\lambda)\beta^{-1/2}$, we
arrive at the desired result, as long as $\beta$ is large enough to ensure $\eps<1$. But
if $\beta\leq (2+\lambda)^2$, (\ref{320}) holds trivially because of the positivity of $\rho$.
\end{proof}

\subsection{The limit of very strong magnetic fields} \label{highfieldlimit}

In the investigation of the limit $\beta \to \infty$ some special
concepts from earlier studies are involved: The right scaling, the
restriction to the lowest Landau band, the effects of the
superharmonicity on the distribution of the density in the
perpendicular directions, and the high field limit of the Coulomb
interaction.

We consider pure states $\Lambda =|\chi \rangle \langle \chi |$
defined by the wave functions with $\Lpar =0$
in the lowest Landau band,
\beq \label{chi}
\chi (\x )=\sqrt{\frac{\beta}{2\pi}}%
e^{-\beta (\xperp)^2 /4}%
\sqrt{L}\psi (L\xpar ).
\eeq
We denote here the perpendicular and parallel components of the
coordinates as $\x = (\xperp , \xpar )$. The scaling factor
$L=L(\beta )$ has been defined in \cite{BSY00} as the solution of
the equation
\beq \beta ^{1/2}=L(\beta )\sinh[L(\beta )/2].
\eeq
Note that
\beq L(\beta )=\ln \beta +O(\ln\ln \beta )\eeq
as $\beta \to \infty$.
In the following we use the scaled coordinates
\beq \label{betascal}
z=L(\beta)\xpar ,\qquad \rr = \sqrt{\beta}\xperp , \qquad r=|\rr |.
\eeq
With the density matrix $\Lambda =|\chi \rangle \langle \chi |$,
we calculate the contributions to $\E_\beta [\Lambda]$. The
normalization of $\Lambda$, we denote it as $\lambda _\psi$, is
equal to $\|\psi\|_2^2$. We restrict the set of $\psi$ to real
valued wave functions, since they minimize the kinetic energy, if
the density $|\psi|^2$ is kept fixed.

\begin{lem}[The strong field limits]\label{highblim}
Given the state $\Lambda$ defined by real valued $\psi \in
\mathcal{H}^{1}(\R)$ as in (\ref{chi}), the contributions to the
energy in the magnetic Hartree theory, magnetic-kinetic energy
$K_{\beta ,\psi}$, attraction $A_{\beta ,\psi}$, and repulsion
$R_{\beta ,\psi}$, are in the following way related to the kinetic
energy, $K_\psi = \int (d\psi / dz)^2 dz$, attraction-energy
$A_\psi =\psi (0)^2$, and repulsion-energy $R_\psi = \frac{1}{2}
\int \psi (z)^4 dz$ of the density $\psi (z)^2$ in the HS-theory:
\beqa
K_{\beta ,\psi}&=& L^2 K_\psi , \label{khighblim} \\
\left|\frac{1}{L^2}A_{\beta ,\psi}-A_\psi \right| &\leq&
\frac{C}{L}\left( \lambda _\psi + \lambda _\psi ^{1/4}K_\psi
^{3/4} \right), \label{ahighblim} \\ \left|\frac{1}{L^2}R_{\beta
,\psi}-R_\psi \right| &\leq& \frac{C\lambda _\psi}{L}\left(
\lambda _\psi + \lambda _\psi ^{1/4}K_\psi ^{3/4} \right).
\label{rhighblim}
\eeqa
\end{lem}

\begin{proof}
The equation for the kinetic energy is obvious by definition of
$\Lambda$. We calculate the energy of attraction as
\beq
A_{\beta ,\psi}= L^2\int_0^{\infty}e^{-r^2/2}
\left[\int_{-\infty}^{\infty}V_{\beta ,r}(z)\psi ^2(z)dz\right]rdr,
\eeq
where
\beq
V_{\beta ,r}(z)=\frac{1}{L\sqrt{L^2r^2/\beta +z^2}}.
\eeq
For the term in square brackets we use Lemma 2.1 of \cite{BSY00},
\beq
\left| [...] - \psi (0)^2 \right| \leq \lambda _\psi /r+8\lambda
_\psi^{1/4}K_\psi ^{3/4}r^{1/2},
\eeq
to estimate the difference to the expected limit in the HS-theory:
\beq
\left|\frac{1}{L^2}A_{\beta ,\psi}-A_\psi \right| \leq
\frac{1}{L}\left( \lambda _\psi \sqrt{\pi /2}+
8\cdot 2^{1/4}\Gamma (5/4)\lambda _\psi^{1/4}K_\psi^{3/4}\right).
\eeq
For the repulsion $R_{\beta ,\psi}$ we calculate
\beq \label{rhighb}
\left|\frac{1}{L^2}R_{\beta ,\psi}-R_\psi \right|     \leq
\frac{1}{2} \int_{\R^4}^{}d^2\rr d^2\rr '(\frac{1}{2\pi})^2
e^{-(\rr ^2+\rr '^2)/2}[...],
\eeq
where
\beqa
[...] =  \left|  \int_{\R^2}dz dz' \left( V_{\beta ,|\rr -\rr
'|}(z-z')-\delta (z-z') \right) \psi^2 (z) \psi^2 (z') \right|
\label{rpsi} \\ \leq \frac{1}{L}\left(\frac{\lambda _\psi ^2}{|\rr
-\rr '|}+ 8\lambda _\psi^{5/4}K_\psi^{3/4}|\rr -\rr
'|^{1/2}\right) . \label{rhighc}
\eeqa
The inequality is supplied by Lemma 2.2 of \cite{BSY00}, with an
adaptation due to the different notation concerning the
normalization. After inserting (\ref{rhighc}) in (\ref{rhighb}),
the integrals can be evaluated as
\beq
\left|\frac{1}{L^2}R_{\beta ,\psi}-R_\psi \right|\leq
\frac{1}{L}\left( (\sqrt{\pi}/4)\lambda _\psi^2 +
4\sqrt{2}\Gamma (5/4)\cdot \lambda _\psi^{5/4}K_\psi^{3/4} \right) .
\eeq

\end{proof}

\begin{thm}[Magnetic Hartree energy for large $\beta$]\label{nrglarge}
In the limit
\newline $\beta \to \infty$
with $\lambda$ fixed,
\beq
\lim_{\beta \to\infty}\frac{E^{\rm MH}(\lambda,\beta)}{(\ln \beta )^2}=
E^{\rm HS}(\lambda).
\eeq
\end{thm}

\begin{proof}
Combining all the bounds of the Lemma, and using $\lambda _\psi
^{1/4}K_\psi ^{3/4}\leq \frac{1}{4}\lambda _\psi
+\frac{3}{4}K_\psi$ to simplify, gives
\beq \label{comparison}
\left|\frac{1}{L^2}\E_{\beta}[\Lambda] -{\mathcal E}^{\rm
HS}[|\psi |^2]\right| \leq \frac{C}{L(\beta)}(1+\lambda _\psi
)(\lambda _\psi+ K_\psi).
\eeq

For the {\bf upper bound} on $E^{\rm MH}(\lambda,\beta)$ we
specify $\psi$ as $\psi _\lambda ^{\rm HS}$, the minimizing wave
function for the HS-theory. See equation (3.6) in \cite{LSY94a} or
the following (\ref{rhohs}). It remains $\psi _2 ^{\rm HS}$, for
all $\lambda > 2$, so these wave functions are normalized to
$\|\psi \|_2^2 = \lambda _\psi = \min \{\lambda ,2\}$. Since the
variational principle with fixed norm can be replaced by a
variational principle with bounded norm, as we have remarked in
the Subsection \ref{hartreedef}, the wave function $\psi _2 ^{\rm
HS}$ is good for the upper bounds for all $\lambda \geq 2$.
$K_\psi$ is the kinetic energy in the hyperstrong theory. Using
the virial equation (\ref{virial}), we may replace $K_\psi $ on
the right side of (\ref{comparison}) by $|E^{\rm HS}(\lambda )|$.
With this choice of $\psi$ the equation (\ref{comparison}) gives
the upper bound
\beq \label{comparisonupper}
\frac{1}{L^2}E^{\rm MH}(\lambda,\beta) \leq  E^{\rm HS}(\lambda) +
\frac{C}{L(\beta)} \left( \lambda+ \left| E^{\rm
HS}(\lambda)\right| \right) .
\eeq

To derive a {\bf lower bound}, we use the error bound, when
confining the theory to the lowest Landau band, which we have
estimated in the Subsection \ref{lowestlandauhartree}. By Lemma
\ref{compconf} we have, for $\beta$ large enough,
\beq \label{compconf2}
E^{\rm MH}(\lambda,\beta) \geq (1-{\rm
const.}(1+\lambda)^2\beta^{-1/2})^{-1} E_{\rm conf}^{\rm
MH}(\lambda,\beta).
\eeq

Also this confined Hartree theory has rotation invariant
minimizers, since the lowest Landau band is mapped onto itself by
rotations around the z-axis. Moreover, since the potential is
superharmonic, also the minimizer of this confined theory has
$\Lpar =0$. See Corollary \ref{conf0} of Subsection \ref{min0}.
The variational principle can therefore be restricted to states
$\Lambda  =|\chi \rangle \langle \chi |$, where the wave functions
$\chi$ are specified as in equation (\ref{chi}), with general
$\psi \in \mathcal{H}^{1}(\R)$, normalized to $\|\psi\|_2^2
=\lambda$:
\beq
E_{\rm conf}^{\rm MH}(\lambda,\beta) =
\inf_{\Lambda , \hspace{4pt} \Tr[\Lambda]=\lambda , \hspace{4pt}
\Lambda  =|\chi \rangle \langle \chi |}
\E_{\beta}[\Lambda].
\eeq

The comparison with HS-theory in (\ref{comparison}) is now used as
a lower bound
\beq
\frac{1}{L(\beta)^2}\E_{\beta}[\Lambda]\geq
{\mathcal E}^{\rm HS}[|\psi |^2]
-\frac{C}{L(\beta)}(1+\lambda )(\lambda + K_\psi ).
\eeq
The right side of this inequality can be considered as a
functional similar to the HS-functional, but with the constant
$1-C(1+\lambda )/L(\beta)$ in front of the kinetic energy. With an
appropriate scale transformation, this is equivalent to a
HS-functional multiplied with $\left( 1-(1+\lambda )C/L(\beta)
\right) ^{-1}$, as long as this parameter is positive. Taking the
infima of both sides, we conclude that
\beq
\frac{1}{L(\beta)^2} E_{\rm conf}^{\rm MH}(\lambda,\beta) \geq
\left( 1-C\frac{(1+\lambda )}{L(\beta)} \right)^{-1} E^{\rm
HS}(\lambda) -C\frac{(1+\lambda )\lambda}{L(\beta)}
\eeq
for $\beta$ large enough. Considering the limit $\beta \to \infty$
at constant $\lambda$ we infer
\beq
\liminf_{\beta\to\infty}  \frac{1}{L(\beta)^2} E_{\rm conf}^{\rm
MH}(\lambda,\beta) \geq E^{\rm HS}(\lambda).
\eeq
Combining this with (\ref{compconf2}) proves, in union with
(\ref{comparisonupper}), the theorem.
\end{proof}

\begin{rem}\label{specify}
To be precise, we specify the asymptotics:
\beqa
\frac{E^{\rm MH}(\lambda,\beta)}{L( \beta )^2} &\leq& \left(
1-C\frac{\lambda}{L(\beta)} \right) E^{\rm HS}(\lambda) +
C\frac{\lambda}{L(\beta)}, \\ \nonumber \frac{E^{\rm MH}(\lambda,\beta)}{L(
\beta)^2} &\geq& \left(1-C\frac{(1+\lambda)^2}{\beta^{1/2}}\right)^{-1}\left(
1-C\frac{1+\lambda}{L(\beta)}\right)^{-1}
E^{\rm HS}(\lambda) - C\frac{\lambda+\lambda^2}{L(\beta)},\\
\eeqa
as long as the terms in braces are positive.
\end{rem}

\begin{rem}
The convergence both of the energy of the atom, and of the energy
per unit charge, when divided by $(\ln \beta)^2$, is uniform in
$\lambda$ on bounded sets of $\lambda$.
\end{rem}

Analyzing the limit $N \to \infty$ of many particle quantum mechanics, we
will be led to the linearized mean field theory, where we need the
\begin{lem}[Generalized strong field limits] \label{highblim2}
Given two states which are defined by real valued $\psi \in
\mathcal{H}^{1}(\R)$ and real valued $\varphi \in
\mathcal{H}^{1}(\R)$ as in (\ref{chi}), with densities
$\rho_{\beta,\psi} = \beta /2\pi e^{-\beta (\xperp)^2 /2} L\psi
(L\xpar )^2$, the following estimate for the Coulomb repulsion
holds:
\beq
\left|\frac{1}{L^2} D[\rho_{\beta,\varphi},\rho_{\beta,\psi}] -
\frac{1}{2}\int_{-\infty}^{+\infty}\varphi (z)^2\psi(z)^2dz\right|
\leq
\frac{C\lambda_\varphi}{L}\left(\lambda_\psi+\lambda_\psi^{1/4}
K_\psi^{3/4}\right).
\eeq
\end{lem}
\begin{proof}
As for $R_{\beta,\psi}$ in the proof of Lemma \ref{highblim}, with
$\psi(z)^2$ in (\ref{rpsi}) changed to $\varphi(z)^2$.
\end{proof}

\section{The mean field limit} \label{highnlimit}

We now prove the theorems that we have stated in the introduction.
We will derive appropriate upper and lower bounds to the quantum
mechanical energy of symmetric states obeying Bose statistics.

The atom with many interacting particles will be compared to
models with independent particles in an effective field. In this
comparison the upper bound is rather easy: One uses symmetric wave
functions of product form as trial wave functions. Then one adds
the ``self energies'' of the one-particle densities.

The production of lower bounds is not so easy. For the Regions $1$
and $2$ we will use the Lieb-Oxford bound, \cite{LO81}, and then
we will borrow a part of the kinetic energy to estimate the
correction in comparison to the mean field model. But this method
breaks down in the presence of magnetic fields, when they are too
strong. In Region $3$ its use is restricted to the subregion
$\beta \lesssim N^{4/3}$. This situation is similar to the case of
fermionic electrons, discussed in Sect.~$7$ of \cite{LSY94a}.

For $\beta \gtrsim N^{4/3}$ we have to borrow some kinetic energy
at the level of quantum mechanics for $N$ particles. Here we
develop a new way of producing bounds, extending the method of
\cite{BSY00}.

\subsection{Upper bounds} \label{nparticleupperbounds}

\begin{lem}[Upper bounds for Regions $1$ and $2$]\label{up1-2}
The Hartree energies provide upper bounds to the quantum
mechanical energies for each $N$, $\lambda$ and $\beta$:
\beq
\frac{E(N,\lambda,\beta)}{N}\leq
\frac{E^{\rm MH}(\lambda,\beta)}{\lambda}.
\eeq
\end{lem}

\begin{proof}

We apply the variational principle, using $N$-fold products of
$1$-particle states $\Gamma$ as $N$-particle states
\beq
E(N,\lambda,\beta)\leq
N \Tr[(H_\beta-\beta)\Gamma]-N\int\frac{1}{|\x|}\rg
+\lambda \frac{N(N-1)}{N^2} D[\rg,\rg]
\eeq
for all $\Gamma$ with $\Tr[\Gamma]\leq 1$. Setting
$\Gamma=\Gamma^{\rm H}/\lambda$ we see that the inequality holds, for
each $\beta$ and $\lambda$.
\end{proof}

\begin{lem}[Upper bound for Region $3$]\label{up3}
In the limit $\beta \to \infty$, the energies of the hyperstrong
theory are asymptotic upper bounds to the quantum mechanical
energies for each $N$:
\beq
\limsup_{\beta \to \infty} \frac{E(N,\lambda,\beta)}{N (\ln \beta)^2}\leq
\frac{E^{\rm HS}(\lambda)}{\lambda}.
\eeq
To be precise:
\beq
\frac{E(N,\lambda,\beta)}{N\,L( \beta )^2} \leq \left(
1-C\frac{\lambda}{L(\beta)} \right) \frac{E^{\rm
HS}(\lambda)}{\lambda} + \frac{C}{L(\beta)}.
\eeq

\end{lem}

\begin{proof}
This results from combining Lemma \ref{up1-2} with Theorem
\ref{nrglarge}.
\end{proof}

\subsection{Lower bounds for Regions $1$ and $2$} \label{nparticlelowerbounds}

\begin{lem}[Lower bounds for Regions $1$ and $2$]\label{down1-2}
In the limit $N \to \infty$, the Hartree energies are
asymptotic lower bounds to the quantum mechanical energies,
\beq
\liminf_{N \to \infty} \frac{E(N,\lambda,\beta)}{N }\geq
\frac{E^{\rm MH}(\lambda ,\beta )}{\lambda},
\eeq
uniformly in $\beta$ for bounded $\beta$.
\end{lem}

\begin{proof}
We use the Lieb-Oxford inequality \cite{LO81}
\beq
\langle\Psi|\sum_{i<j}|\x_i-\x_j|^{-1}|\Psi\rangle\geq
D[\rho_\Psi,\rho_\Psi]-C\int\rho_\Psi^{4/3}
\eeq
($\Psi$ is normalized as $\|\Psi\|_2 =1$, the constant can be
chosen to be $C=1.68$), which, together with H\"older's inequality
$\int\rho^{4/3}\leq (\int\rho^3)^{1/6}(\int\rho)^{5/6}$, implies
that
\beq\label{low}
\langle\Psi|H_{N,\lambda,\beta}\Psi\rangle\geq\frac N\lambda
\E_\beta [\Gamma_\Psi\lambda/N]-C\lambda N^{-1/6}\left(
\int\rho_\Psi^{3}\right)^{1/6},
\eeq
where $\Gamma_\Psi$ is the one-particle reduced density matrix of
$\Psi$, and $\rho_\Psi$ is its density. Now if
$\langle\Psi|H_{N,\lambda,\beta}\Psi\rangle\leq 0$, which we can
of course assume, then
\beq\label{eps}
\langle\Psi|\sum_i(H_{\beta,i}-\beta)\Psi\rangle\leq
-\langle\Psi|\sum_i(H_{\beta,i}-
\beta-2|\x_i|^{-1})\Psi\rangle\leq -N E^{\rm hyd}(\beta,2),
\eeq
with $E^{\rm hyd}$ defined in (\ref{hydro1}). Together with
(\ref{l3}) this implies that
\beq
\left(\int\rho_\Psi^3\right)^{1/3}\leq \frac13\left(\frac
2\pi\right)^{4/3} N\left(\beta-E^{\rm hyd}(\beta,2)\right),
\eeq
so finally
\beq\label{loweps}
\frac1N E(N,\lambda,\beta)\geq \frac1\lambda E^{\rm
MH}(\lambda,\beta)- C\lambda N^{-2/3}\left(\beta-E^{\rm
hyd}(\beta,2)\right)^{1/2}.
\eeq
\end{proof}

\begin{rem}
Note that the convergence of the energies in Region 2, including
Region 1, is uniform in $\beta$ for bounded $\beta$. So if
$\beta\to 0$ as $N\to\infty$ we get the usual Hartree energy
without magnetic field. It is even possible to let
$\beta\to\infty$ with $N$ as long as $\beta\leq {\rm const.}
N^{4/3}$. In fact, in (\ref{loweps}) we have an error term of
order $N^{-2/3}\beta^{1/2}$ (note that $E^{\rm hyd}\sim
(\ln\beta)^2$ for large $\beta$), and this is of lower order than
$E^{\rm MH}$ as long as $\beta\leq {\rm const.} N^{4/3}$.
\end{rem}

\subsection{Restriction to independent particles}\label{restindependent}

An essential ingredient is the positive definiteness of the
repulsive pair interaction. The method exploiting positive
definiteness of a function $W$ uses the inequality
\begin{eqnarray}\label{pdef}
\sum_{i<j}W(\x_{i}-\x_{j}) &\geq &\sum_{i=1}^{N}\int_{-\infty
}^{+\infty }W(\x_{i}-\y)\sigma (\y)d^3\y  \\ && -
\frac{1}{2}\int\!\!\!\int \sigma (\x)W(\x-\y)\sigma (\y)d^3\x
\,d^3\y -  N\frac{W(\0)}{2}. \nonumber
\end{eqnarray}
Actually we need a function $W$ which is positive definite, finite
at the origin, and a lower bound to the Coulombic repulsive
potential. Also the cutoff near the origin should vanish in the
limits $N \to\infty$, $\beta\to\infty$, but still with $W(\0
)/N(\ln\beta)^2$ going to $0$. If we choose for $W$ the spherical
symmetric cutoff potential
\beq
V_{\rm cutoff}(\x)=\frac{1-e^{-\mu |\x |}}{|\x|},
\eeq
it turns out that this does not suffice for our purposes. The
cutoff length $\mu^{-1}$ should be of the order of the typical
lengths of the atom, and this would require a coupling of the
limits $N\to\infty$ and $\beta\to\infty$, or more precisely,
$\beta$ is not allowed to increase arbitrarily fast with $N$.

To get a useful lower bound for the entire Region 3, we have to
push down $W(\0)$ even further, without changing the effective
interaction to much. To achieve this, we split $V_{\rm cutoff}$
(we restrict the parameter to $\mu>1$) into its effective part
$V_\mu$ and the long range tail $V_{\rm long}$,
\beqa
V_\mu (\x) &=& \frac{e^{-|\x|}-e^{-\mu |\x|}}{|\x|},\qquad \mu >1, \\
V_{\rm long}(\x) &=& \frac{1-e^{-|\x|}}{|\x|},
\eeqa
and borrow some kinetic energy.

The {\bf construction of ${\bf W}$} proceeds in several steps.
We begin with the observation that
$V_\mu$ is integrable in $\xpar$ for any $\xperp$.
It has moreover the property of being decreasing in $|\x|$. So
the integrals along lines with fixed $\xperp$ can be bounded by
the integral along the line, where $\xperp = 0$:
\beq
\int_{-\infty}^\infty V_\mu (\x) d\xpar \leq 2 \ln \mu.
\eeq

We now use Lemma $6.3$ of \cite{BSY00}, the {\bf operator inequality},
which holds for $\alpha>0$, $F>0$, and for each $b>0$,
$b$ independent of $\alpha$ and $F$,
\beq \label{deltiq}
\alpha p_x^2  +F\delta (x) \geq F\cdot w_{F/\alpha , b}(x),
\eeq
where
\beq \label{deltappr}
w_{D, b}(x) = D\frac{b^2}{2b+1}e^{-bD|x|}
\eeq
and $p_x^2=-(\partial/\partial x)^2$. These functions have the
properties of being positive definite and producing delta-function
sequences in the limit $bD \to \infty$ coupled with $b \to
\infty$. (The variable $x$ will in the application be replaced by
$\xpar $.) We extend this lemma to the
\begin{proposition}[Operator inequalities]
Let V(x) define a positive Borel measure $V(x)dx$,
with $\int_{-\infty }^{+\infty }V(x)dx \leq F$,
$F>0$, $\alpha>0$.
For each $b>0$ the operator inequality
\beq
\alpha p_x^2  +V(x)\geq (V * w_{F/\alpha , b})(x) \label{opiq4}
\eeq
holds, in the sense of an inequality for quadratic forms, with the
Sobolev space ${\mathcal H}^1 (\R)$ as the form domain.
\end{proposition}

\begin{proof}
For $\psi \in {\mathcal H}^1 (\R)$ consider the wave functions
$\psi_y(x) = \psi (x+y)$.
By (\ref{deltiq}) there holds for each $y$ the inequality
\beq
\langle \psi_y | ( \alpha p^2 +F\delta ) |\psi_y \rangle%
\geq F\langle \psi_y | w_{F/\alpha , b} |\psi_y \rangle.
\label{functiq}
\eeq
Both sides of this inequality are Borel measurable functions of $y$.
Observe $\langle \psi_y | p^2 |\psi_y \rangle =%
\langle \psi | p^2 |\psi \rangle$,
$\langle \psi_y | w(x) |\psi_y \rangle =%
\langle \psi | w(x-y) |\psi \rangle$,
and integrate both sides of (\ref{functiq}) with $V(y)dy$.
Divide by $F$, and use $\int_{-\infty }^{+\infty }V(x)dx / F \leq 1$
in front of the kinetic energy term.
\end{proof}

We may consider $\alpha (\ppar_i)^2 + V_{\mu }(\x_{i}-\x_{j})$ as
a 1-particle Hamiltonian acting in $\mathcal{L}^{2}(\R
,d\xpar_{i})$, which is parametrized by the perpendicular
coordinates and by $\x_{j}$. However, the operator inequality
(\ref{opiq4}) is also valid if the operators act in the extended
space $\mathcal{L}^{2}(\R ^{3N},d^3 \x_{1}...d^3 \x_{N})$, so we
can apply this proposition, with $F=2\ln \mu$, to
\beq
\alpha (\ppar _i)^2 + V_{\mu }(\x_i-\x_j).
\eeq

We add the missing long range tail and define
\beq \label{wdef}
W(\x)=\int V_{\mu }(\x-\y) \delta ^2 (\yperp ) w_{2\ln \mu /\alpha
,b}(\ypar)d^3\y +\frac{1-e^{-|\x|}}{|\x|},
\eeq
where $w_{...}(\xpar)$ is defined in (\ref{deltappr}).
The essential effect of the convolution of $V_\mu$ with
the distribution
\beq
\uw (\x)=\delta(\xperp)w(\xpar)
\eeq
is the lowering of the value of the interactions at the points of
coincidences of particles,
\beq
W(\0) \leq \left( \int V_\mu d\xpar \right) w_{2\ln \mu /\alpha ,b}(0) +1
< \frac{2b}{\alpha}(\ln \mu)^2 +1.
\eeq

Thus we have shown, that $W(\x)$ is a lower bound to the Coulomb
potential, when some borrowed part of the kinetic energy is added,
\beq
\label{wleqc}
W=V_\mu *\uw + V_{\rm long} \leq \alpha (\ppar)^2 + V_\mu +V_{\rm long}
\leq \alpha (\ppar)^2 +1/|\x|.
\eeq
Now the function $W = V_\mu *\uw +V_{\rm long}$ is also positive definite,
since the distribution $\uw$ and all the involved functions
are positive definite.
So we can refer to the technique (\ref{pdef})
of reduction to one-particle models.

\begin{proposition}[Bounds by independent particles]\label{bdbyind}
Considered as quadratic forms, the N-particle Hamiltonians
$H_{N,\lambda ,\beta }$ are bounded from below by sums of
one-particle Hamiltonians and constants:
\beq
H_{N,\lambda ,\beta }  \geq  \sum_{i=1}^{N}h_{i} - \frac12
\frac{\lambda}{N} \int\!\!\!\int \sigma (\x )W(\x -\y)\sigma
(\y)d^3\x \,d^3\y -\lambda \frac{b}{\alpha}(\ln \mu
)^{2}-\frac{\lambda}{2},
\eeq
where
\beq \label{oneham}
h_{i}= H_{\beta , i} - \beta -\frac{\alpha \lambda}{2}(\ppar _i)^2
- \frac{1}{|\x_{i}|}+\frac{\lambda}{N}\int W(\x _{i}-\y )\sigma
(\y )d^3\y,
\eeq
and where
$W(\x)$ is defined in (\ref{wdef}).
The parameters $\alpha ,b$ have to be positive, $\mu>1$,
$\sigma (\x)$ should be an element of $\Ll^1 (\R^3)$.
\end{proposition}

\begin{proof}
From $H_{\beta ,i}$ we borrow a part of $(\ppar_i)^2 = -
(\partial /\partial \xpar_i)^2$
to use (\ref{wleqc}),
\beq
\alpha (\ppar _i)^2 + \frac{1}{|\x_i-\x_j|} \geq  W(\x_i-\x_j),
\eeq
sum over all pairs $i\neq j$ , and multiply with $\lambda /2N$.
This gives
\beq \label{opiq3}
\alpha \lambda \frac{N-1}{2N}\sum_{i=1}^{N}(\ppar _i)^2
+\frac{\lambda }{N} \sum_{i<j}\frac{1}{|\x_i-\x_j|} \geq
\frac{\lambda }{N}\sum_{i<j}W(\x _{i}-\x _{j}).
\eeq
Inserting (\ref{opiq3}) and (\ref{pdef}) into (\ref{ham2})
completes the proof.
\end{proof}

To ascertain any use to the inequality, $\alpha \lambda <2$ is
requested. In the next subsection, the density $\sigma$ will be
chosen to be an approximately minimizing Hartree density, and the
parameters will be adapted to the limit $N,\beta \to \infty $.

\subsection{The lower bound for Region $3$}\label{reg3}

We choose $\sigma = N\rho$, with
\beq
\rho(\x) = \frac{\beta}{2\pi} \exp \left( -\frac{\beta (\xperp
)^2}{2} \right) \frac{1}{\lambda}L\rho_\lambda^{\rm HS}(L\xpar ).
\eeq
The minimizing density  $\rho_\lambda^{\rm HS}$ for the
hyperstrong theory is, see \cite{LSY94a},
\beqa \label{rhohs}
\rho_\lambda^{\rm HS}(z) &=& \frac{2(2-\lambda)^2} {(4\sinh
[(2-\lambda )|z|/4+c(\lambda)])^2} \qquad {\rm for} \quad \lambda
<2, \\ && \qquad \tanh c(\lambda) = (2-\lambda )/2,  \nonumber \\
\rho_\lambda^{\rm HS}(z) &=& 2(2+|z|)^{-2} \qquad \qquad \qquad
\qquad\quad\ \ {\rm for} \quad \lambda \geq 2.
\eeqa
Note that $\|\rho\|_1 = \min\{1,2/\lambda\}, \quad \|\rho\|_\infty
= (\beta /2\pi )L (\beta)\rho_\lambda^{\rm HS}(0)/\lambda \leq
C\beta L(\beta)$.

\bigskip

We want to study the lower bound to $N^{-1}E(N,\lambda,\beta)$,
due to Proposition \ref{bdbyind},
\beq
\infspec \{h\}-\frac{\lambda}{2} \int\!\!\!\int \rho (\x ) W(\x
-\y)\rho (\y)d^3\x \,d^3\y -\frac{\lambda}{N}\left(
\frac{b}{\alpha}(\ln \mu )^{2}+\frac 12\right),
\eeq
which we expect to be asymptotically proportional to $(\ln
\beta)^2$. For simplicity, we will restrict our considerations to
$\beta>C_\beta$ for some $C_\beta>1$ in the following. We choose
the parameters as
\beq
\mu=\beta^{1/2+\delta}, \alpha=N^{-\eps}, b=N^\eta ,
\eeq
with $\delta, \eps, \eta$ all greater than zero, $\eps+\eta<1$.
This guarantees first of all the relative vanishing of the
constant which stems from $W(\0)$:
\beq
\frac{\lambda}{N} \left(\frac{b}{\alpha}(\ln \mu )^{2}+\frac 12\right)
/(\ln \beta )^2 \to 0,
\eeq
as $\beta,N \to \infty$. It remains to study the Hamiltonian $h$
and the self energy of $\rho$ due to the interaction $W=V_\mu *\uw
+ V_{\rm long}$.

The inequality
\beq
\int \!\!\!\int \rho(\x)(V_\mu *\uw )(\x -\y) \rho (\y)d^3\x \,d^3\y \leq
\int \!\!\!\int \rho(\x)V_\mu(\x -\y) \rho (\y)d^3\x \,d^3\y
\eeq
can be observed in Fourier space, where $\widetilde V_\mu$ and
$\widetilde \uw$ are positive, and $\widetilde \uw \leq 1$. We add
$V_{\rm long}$, and use the pointwise inequality in $\x$-space
$(V_\mu +V_{\rm long})(\x) \leq V^C(\x)\equiv 1/|\x|$. So the
self-energy of $\rho$ due to $W$ is smaller than the self-energy
$D[\rho,\rho]$ due to the interaction by the Coulomb potential
$V^C$.

The potential $(V_\mu *\uw +V_{\rm long})*\rho$ in the Hamiltonian
$h$ is now changed to $V^C*\rho$. First we give a bound to the
difference of $V_{\mu}*\uw *\rho$ to $V_{\mu}*\rho$ in the form of
an operator inequality, which follows from
\beqa
|\,\langle \psi| V_{\mu}*(\uw *\rho -\rho )|\psi\rangle\,| &\leq&
\|\,(|\psi|^2*V_{\mu})\cdot(\uw *\rho -\rho)\|_1  \nonumber \\
&\leq& \|\uw *\rho -\rho\|_1\,\,\|\,|\psi|^2*V_{\mu}\|_\infty
\nonumber  \\ &=& \|\uw *\rho
-\rho\|_1\,\sup_{\y}\langle\psi|V_{\mu,\y}|\psi\rangle, \label{vy}
\eeqa
where $V_{\mu,\y}=V_{\mu}(\x -\y)$. Since $V_{\mu}\leq V^C$, and,
with $E^{\rm hyd}(\beta)\equiv E^{\rm hyd}(\beta,\zeta=1)$,
\beq
H_\beta -\beta - V_{\y}^C \geq E^{\rm hyd}(\beta),
\eeq
the last term in (\ref{vy}) is bounded by
\beq
\langle\psi|V_{\y}^C|\psi\rangle \leq \langle\psi|
H_\beta -\beta - E^{\rm hyd}(\beta)|\psi\rangle,
\eeq
multiplied with $\|\uw *\rho -\rho\|_1$.

We observe, see Remark \ref{ll0},
\beq
E^{\rm hyd}(\beta)=\lim_{\lambda\to 0}\frac{E^{\rm
MH}(\lambda,\beta)}{\lambda}, \qquad \lim_{\lambda\to
0}\frac{E^{\rm HS}(\lambda)}{\lambda}=-\frac{1}{4}.
\eeq
The precise bound in the Remark \ref{specify} gives thus (see also
\cite{AHS81})
\beq
E^{\rm hyd}(\beta) \geq \left(
-\frac{1}{4}-C\frac{1}{L(\beta)}\right) L^2(\beta)\geq
-C(\ln\beta)^2
\eeq
for $\beta>C_\beta$. An analogous bound holds also for $E^{\rm
hyd}(\beta,2)$. Note that $L(\beta)$ can be replaced by
$(\ln\beta)$ (and vice versa), since for $\beta>C_\beta>1$ there are positive
constants $C_1$ and $C_2$ such that
$C_1\leq\ln(\beta)/L(\beta)\leq C_2$.

To complete the estimate (\ref{vy}), we need a bound to $\|\uw
*\rho -\rho\|_1$. There, the integrals in perpendicular
coordinates can be done explicitly. Note that for all $\lambda$ we
have $|d\rho^{\rm HS}_\lambda/d z|\leq  \rho^{\rm HS}_\lambda$. Using this and the
monotonicity of $\rho^{\rm HS}_\lambda$ in $|z|$ we can estimate
\begin{eqnarray}\nonumber
\left|\rho^{\rm HS}_\lambda(z-y)-\rho^{\rm HS}_\lambda(z)\right|&\leq& |y|
\sup_{x\in[z-y,z]}\left|\frac{d \rho^{\rm HS}_\lambda}{dz}(x)\right|
\\ \nonumber &\leq& |y|\left(
\rho^{\rm HS}_\lambda(z-y)+\rho^{\rm HS}_\lambda(z)+\Theta(|y|-|z|)
\rho^{\rm HS}_\lambda(0)
\right).\\ \label{Theta}
\end{eqnarray}
Since for $|y|\geq 1$ (\ref{Theta}) is obviously true even without
the last term, we see that (\ref{Theta}) holds with
$\Theta(|y|-|z|)$ replaced by $\Theta(1-|z|)$. This implies
\begin{eqnarray}\nonumber
\|\rho*\uw -\rho\|_1&\leq& \frac 1\lambda\int d\xpar d\eta L
w(\eta)\\ \nonumber & &\times\left( \left|\rho^{\rm HS}_\lambda(L\xpar
-L\eta)-\rho^{\rm HS}_\lambda(L\xpar )\right| +\left((\mbox{$\int
w$})^{-1}-1\right)\rho^{\rm HS}_\lambda(L\xpar )\right)
\\ \nonumber &\leq& \lambda^{-1}\mbox{$\int\rho^{\rm HS}_\lambda$}(1-\mbox{$\int w$})+
2\lambda^{-1}\left(\mbox{$\int\rho^{\rm HS}_\lambda$}+
\rho^{\rm HS}_\lambda(0)\right) L \int |\eta|w(\eta)d\eta.\\
\end{eqnarray}
Now $\int\rho^{\rm HS}_\lambda\leq \lambda$, $\rho^{\rm HS}_\lambda(0)\leq
\half\lambda$, $(1-\int w)\leq (2b)^{-1}$ and $\int |\eta|w\leq
\alpha /(2b\ln \mu)$, so we get
\beq
\|\rho*\uw -\rho\|_1\leq \frac1{2b}+ \frac{3}{2}\frac{\alpha
L(\beta)}{b\ln \mu} \leq C(N^{-\eta} + N^{-\eta-\eps} L(\beta)/\ln
\beta )\leq C' N^{-\eta}.
\eeq

\bigskip

Finally, we add to $V_\mu$ the short-range term
\beq
V_{\rm short}(\x)=\frac{e^{-\mu |\x|}}{|\x|},
\eeq
and we find the pointwise bound to $(V^C-V_\mu)*\rho$ as
\beq
\|V_{\rm short}\|_1 \|\rho\|_\infty \leq C\mu^{-2} \beta L(\beta)=
C\beta^{-2\delta}L(\beta).
\eeq

\bigskip

We collect the bounds we have obtained so far:
\beq \label{collbounds}
\frac{E(N,\lambda,\beta)}{N} \geq \infspec \{h_b\} -\lambda
D[\rho,\rho]-C\lambda(\ln \beta)^2\left(N^{\eps+\eta
-1}+N^{-\eta}+ \beta^{-2\delta}L(\beta)^{-1}\right),
\eeq
with the one-particle Hamiltonian
\beq \label{cllh}
h_b = \left( H_{\beta} -\beta \right)\left( 1-C\lambda
N^{-\eta}-2\lambda N^{-\eps}\right) - V^C +\lambda V^C*\rho,
\eeq
where we have used $(\ppar)^2\leq H_\beta-\beta$. We now use
(\ref{eps}) to estimate the error terms in the kinetic energy.
Then we apply the statements of the subsection on the confinement
in the mean field theory, Lemma \ref{confmean}, to estimate
\begin{eqnarray}\nonumber
\frac{E(N,\lambda,\beta)}{N} &\geq& \infspec\left\{\Pi_0\left(
H_\beta-\beta-V^C+\lambda V^C*\rho\right)\Pi_0\right\} -\lambda
D[\rho,\rho] \\ \nonumber
&&-C\lambda(\ln\beta)^2\left(N^{-\eta}+N^{-\eps}+N^{\eps+\eta-1}
+\frac{\beta^{-2\delta}}{L(\beta)}+\frac{2+\lambda}\lambda
\beta^{-1/2}\right).\\
\end{eqnarray}
Corollary \ref{conf0} states that the minimizer of the operator in
question has $\Lpar =0$. The search for the infimum of the
spectrum can therefore be restricted to the use of the same type
of wave-functions as in the investigations on the limit of very
strong magnetic fields in Subsection \ref{highfieldlimit}, Lemma
\ref{highblim} and \ref{highblim2}, and these investigations can
be applied, too. We choose $\eps=\eta=1/3$ and get the lower bound
\beqa\nonumber
&&\frac{E(N,\lambda,\beta)}{N\,L^2(\beta)}  \\ \nonumber &&\geq
\left( \infspec \left\{p_z^2-\delta (z) + \rho^{\rm HS}_\lambda
(z) \right\} -\frac1{2\lambda} \int (\rho^{\rm HS}_\lambda (z))^2
dz \right)  \left( 1-C\frac{1+\lambda}{L(\beta)}\right)^{-1}
\\ \label{bbbound} &&\quad -\ C\left(\lambda
N^{-1/3}+L(\beta)^{-1}\left(\lambda
\beta^{-2\delta}+1+\lambda\right) +(2+\lambda)\beta^{-1/2}\right).
\eeqa
The operator in curly brackets is the Hamiltonian for the
linearized HS theory, with ground state energy $(E^{\rm
HS}+\half\int(\rho^{\rm HS}_\lambda)^2)/\lambda$ (cf. eq.
(\ref{mu})). So our final result can be stated as

\begin{lem}[Lower bound for Region 3]
If $L(\beta)>C(1+\lambda)$, then
\beq \label{bbbbound}
\frac{E(N,\lambda,\beta)}{N\,L^2(\beta)} \geq \frac{E^{\rm
HS}(\lambda)}{\lambda}\left(
1-C\frac{1+\lambda}{L(\beta)}\right)^{-1}-C \left(\lambda
N^{-1/3}+\frac{1+\lambda}{L(\beta)}\right).
\eeq
\end{lem}

Hence the convergence of the lower bound, in the limit $N\to
\infty$, $\beta \to \infty$, is proven.

\begin{rem}
The limits $N\to \infty$ and $\beta \to \infty$ can be considered
independently of each other,
in any succession or combination.
\end{rem}

\subsection{Critical particle number}\label{critpart}
Recall that $\underline E(N,Z,B)$ denotes the
ground state energy of the unscaled
Hamiltonian (\ref{ham}).
\begin{lem}[Limit of the critical charge]\label{critlem}
Define
\beq
N_c(Z,B)=\max\{N|\underline E(N,Z,B)<\underline E(N-1,Z,B)\}.
\eeq
We have
\beq\label{limnc}
\liminf_{Z\to\infty}\frac{N_c(Z,\beta Z^2)}{Z}\geq
\lambda_c(\beta).
\eeq
\end{lem}
\begin{proof}
This follows from the convergence of the energies, by analogous
arguments as in \cite{BL83} or \cite{LSY94a}.
\end{proof}

\begin{rem}\label{critrem}
In \cite{S00}, Theorem 3, the following upper bound on $N_c$ is proven:
\beq\label{nc}
N_c<2Z+1+\frac Z2 \min\left\{\left(1+\frac B{Z^2}\right),
C\left(1+\left[\ln\left(\frac B{Z^2}\right)\right]^2\right)\right\}
\eeq
for some constant $C$ independent of $B$ and $Z$. Inserting (\ref{nc})
into (\ref{limnc})
this proves Lemma \ref{critup}.
\end{rem}

\subsection{Convergence of the density matrices}\label{convdens}

\begin{lem}[Ground states are Hartree minimizers]\label{statesmin}
For fixed $\lambda$ and $\beta$, let $\Psi_N$ be an
$\eps$-approximate ground state of $H_{N,\lambda,\beta}$ defined
in (\ref{ham2}), i.e.
\beq
\langle\Psi_N,H_{N,\lambda,\beta}\Psi_N\rangle\leq
E(N,\lambda,\beta)+\eps(N)
\eeq
for all $N$, with $0<\eps=o(N)$. Let $\Gamma_N$ be the
corresponding one-particle reduced density matrix. Then
$\lambda\Gamma_N/N$ is a minimizing sequence for $\E _\beta $, i.e.
\beq
\lim_{N\to\infty}\E _\beta [\lambda\Gamma_N/N]=E^{\rm MH}(\lambda,\beta).
\eeq
\end{lem}
\begin{proof}
Let $\rho_N$ be the density of $\Gamma_N$. We have
\begin{eqnarray}\nonumber
E^{\rm MH}&\leq&\E _\beta [\lambda\Gamma_N/N]=\frac\lambda N
\Tr[(H_\beta-\beta-|\x|^{-1})\Gamma_N]+\frac{\lambda^2}{N^2}
D[\rho_N,\rho_N]\\ \nonumber &=&\frac\lambda
N\langle\Psi_N|H_{N,\lambda,\beta}\Psi_N\rangle-\frac{\lambda^2}{N^2}
\langle\Psi_N|\sum_{i<j}|\x_i-\x_j|^{-1}\Psi_N\rangle
+\frac{\lambda^2}{N^2}D[\rho_N,\rho_N]\\ \label{minseq} &\leq&
\frac\lambda
N(E(N,\lambda,\beta)+\eps)+\frac{\lambda^2}{N^2}C\int\rho_N^{4/3},
\end{eqnarray}
where we used again the Lieb-Oxford inequality for the last step.
By an analogous argument as in Lemma \ref{down1-2} we see
that the right side of (\ref{minseq}) converges to $E^{\rm MH}$ as
$N\to\infty$.
\end{proof}

Since $\lambda\Gamma_N/N$ is a minimizing sequence for $\E _\beta
$, and $\Gamma^{\rm H}$ is unique, we know that
$\lambda\Gamma_N/N\rightharpoonup\Gamma^{\rm H}$ in weak operator sense.
If $\lambda\leq \lambda_c$, there is even norm convergence. To
show this we need the following general lemma:

\begin{lem}[Weak convergence implies norm
convergence]\label{normc} \quad\, Let $a_N$ be a sequence of
positive trace class operators on a Hilbert space. Suppose that
$a_N\rightharpoonup a$ in weak operator sense, for some positive
trace class operator $a$. If \newline
$\lim_{N\to\infty}\Tr[a_N]=\Tr[a]$, then $a_N\to a$ in trace norm,
i.e.
\beq
\|a_N-a\|_1\equiv \Tr[|a_N-a|]\to 0\quad {\rm as}\quad N\to\infty.
\eeq
\end{lem}
\begin{proof}
See \cite{S79}, Thm. 2.20.
\end{proof}

\begin{thm}[Convergence of the density matrices]\label{densconv}
Let $\Gamma_N$ be as in Lemma \ref{statesmin}, and let
$\lambda\leq\lambda_c$. Then for each fixed $\beta$
\beq
\lim_{N\to\infty}\|\lambda\Gamma_N/N-\Gamma^{\rm H}\|_1=0.
\eeq
\end{thm}
\begin{proof}
Since $\Gamma^{\rm H}$ is unique, $\lambda\Gamma_N/N$ converges weakly
to $\Gamma^{\rm H}$ by Lemma \ref{statesmin} and the proof of Theorem
\ref{exist}. Apply Lemma \ref{normc} to $a_N=\lambda\Gamma_N/N$
and $a=\Gamma^{\rm H}$, and note that $\Tr[\lambda\Gamma_N/N]=\lambda$
and $\Tr[\Gamma^{\rm H}]=\min\{\lambda,\lambda_c\}$.
\end{proof}

\subsection{Bose condensation}\label{bosecond}

Bose condensation, as defined in \cite{PO56}, means that the
largest eigenvalue of the one-particle reduced density matrix of
the ground state for the $N$-particle Hamiltonian is of order $N$
as $N\to\infty$, or equivalently, if
$\liminf_{N\to\infty}\|\Gamma_N\|/\Tr[\Gamma_N]>0$. This is shown
to be the case for our model.

\begin{thm}[Bose condensation in the mean field limit]\label{bec}
Let $\Gamma_N$ be as in Lemma \ref{statesmin}. Then for each fixed
$\beta$ and $\lambda$
\beq
\lim_{N\to\infty}\frac{\|\Gamma_N\|}{\Tr[\Gamma_N]}=
\frac{\min\{\lambda,\lambda_c\}}{\lambda}.
\eeq
In particular, if $\lambda\leq \lambda_c$, and if $\varphi_N$
denotes the normalized eigenvector corresponding to the largest
eigenvalue of $\Gamma_N$ with appropriate phase-factor, we have
\beq
\lim_{N\to\infty}\frac{\|\Gamma_N\|}{\Tr[\Gamma_N]}=1,\quad
\lim_{N\to\infty}\| \varphi_N-\varphi_{\rm H}\|_2=0,
\eeq
where $\varphi_{\rm H}=\lambda^{-1/2}\sqrt{\rho^{\rm H}}$ is the normalized
ground state of $H^{\rm H}$.
\end{thm}

\begin{proof}
We use the notations of the proof of Theorem \ref{densconv}. The
first assertion is easily proved, using that
\beq
\left|\|a_N\|-\|a\|\right|\leq \|a_N-a\|\leq \|a_N-a\|_1
\eeq
and $\|a\|=\min\{\lambda,\lambda_c\}$. To prove the second we
denote $P=|\varphi_{\rm H}\rangle\langle\varphi_{\rm H}|$ and
$P_N=|\varphi_N\rangle\langle\varphi_N|$, and compute
\begin{eqnarray}\nonumber
\Tr[a_NP]&=&\|a_N\|\langle\varphi_N|P\varphi_N\rangle+
\Tr[a_N^{1/2}Pa_N^{1 /2}(1-P_N)]\\
&\leq&\|a_N\|\langle\varphi_N|P\varphi_N\rangle+\Tr[a_N]-\|a_N\|,
\end{eqnarray}
where we have used $P\leq 1$ in the last step.
Therefore $\langle\varphi_N|P\varphi_N\rangle\to 1$ as
$N\to\infty$, which gives immediately the desired result.
\end{proof}

\section*{Acknowledgements}
We thank Jakob Yngvason for continuing interest and fruitful
discussions.

\end{document}